\documentclass[12pt,preprint]{aastex}

\shorttitle{Low obliquities in two multiplanet systems,
revealed through asteroseismology}\shortauthors{Chaplin et al.}

\begin{document}

\title{Asteroseismic determination of obliquities of the exoplanet
  systems Kepler-50 and Kepler-65}

\author{
   W.~J.~Chaplin\altaffilmark{1},
   R.~Sanchis-Ojeda\altaffilmark{2},
   T.~L.~Campante\altaffilmark{1},
   R.~Handberg\altaffilmark{3},
   D.~Stello\altaffilmark{4},
   J.~N.~Winn\altaffilmark{2},
   S.~Basu\altaffilmark{5},
   J.~Christensen-Dalsgaard\altaffilmark{3},
   G.~R.~Davies\altaffilmark{1},
   T.~S.~Metcalfe\altaffilmark{6},
   L.~A.~Buchhave\altaffilmark{7,8},
   D.~A.~Fischer\altaffilmark{5},
   T.~R.~Bedding\altaffilmark{4},
   W.~D.~Cochran\altaffilmark{9},
   Y.~Elsworth\altaffilmark{1},
   R.~L.~Gilliland\altaffilmark{10},
   S.~Hekker\altaffilmark{11,1},
   D.~Huber\altaffilmark{12},
   H.~Isaacson\altaffilmark{13},
   C.~Karoff\altaffilmark{3},
   S.~D.~Kawaler\altaffilmark{14},
   H.~Kjeldsen\altaffilmark{3},
   D.~W.~Latham\altaffilmark{15},
   M.~N.~Lund\altaffilmark{3},
   M.~Lundkvist\altaffilmark{3},
   G.~W.~Marcy\altaffilmark{13},
   A.~Miglio\altaffilmark{1},
   T.~Barclay\altaffilmark{12,16},
   J.~J.~Lissauer\altaffilmark{12}
}

\altaffiltext{1}{School of Physics and Astronomy, University of
  Birmingham, Edgbaston, Birmingham, B15 2TT, UK}

\altaffiltext{2}{Department of Physics, Massachusetts Institute of
  Technology, 77 Massachusetts Avenue, Cambridge, Massachusetts 02139,
  USA}

\altaffiltext{3}{Stellar Astrophysics Centre (SAC), Department of
  Physics and Astronomy, Aarhus University, Ny Munkegade 120, DK-8000
  Aarhus C, Denmark}

\altaffiltext{4}{Sydney Institute for Astronomy, School of Physics,
  University of Sydney, Sydney, Australia}

\altaffiltext{5}{Department and Astronomy, Yale University, New Haven,
  CT, 06520, USA}

\altaffiltext{6}{White Dwarf Research Corporation, Boulder, CO,
  80301, USA}

\altaffiltext{7}{Niels Bohr Institute, Copenhagen University, DK-2100
Copenhagen, Denmark}

\altaffiltext{8}{Centre for Star and Planet Formation, Natural History
  Museum of Denmark, University of Copenhagen, DK-1530 Copenhagen,
  Denmark}

\altaffiltext{9}{McDonald Observatory, The University of Texas,
  Austin, TX 78712, USA}

\altaffiltext{10}{Center for Exoplanets and Habitable Worlds, The
  Pennsylvania State University, University Park, PA, 16802, USA}

\altaffiltext{11}{Astronomical Institute, ``Anton Pannekoek'',
  University of Amsterdam, The Netherlands}

\altaffiltext{12}{NASA Ames Research Center, MS 244-30, Moffett Field,
  CA 94035, USA}

\altaffiltext{13}{Department of Astronomy, University of California,
  Berkeley, CA 94720, USA}

\altaffiltext{14}{Department of Physics and Astronomy, Iowa State
  University, Ames, IA, 50011, USA}

\altaffiltext{15}{Harvard-Smithsonian Center for Astrophysics,
  Cambridge, Massachusetts 02138, USA}

\altaffiltext{16}{Bay Area Environmental Research Institute, 596 First
  St West, Sonoma, CA 95476, USA}

\begin{abstract}

Results on the obliquity of exoplanet host stars -- the angle between
the stellar spin axis and the planetary orbital axis -- provide
important diagnostic information for theories describing planetary
formation.  Here we present the first application of asteroseismology
to the problem of stellar obliquity determination in systems with
transiting planets and Sun-like host stars.  We consider two systems
observed by the NASA \emph{Kepler} Mission which have multiple
transiting small (super-Earth sized) planets: the previously reported
Kepler-50 and a new system, Kepler-65, whose planets we validate in
this paper.  Both stars show rich spectra of solar-like
oscillations. From the asteroseismic analysis we find that each host
has its rotation axis nearly perpendicular to the line of sight with
the sines of the angles constrained at the $1\sigma$ level to lie
above 0.97 and 0.91, respectively. We use statistical arguments to
show that coplanar orbits are favoured in both systems, and that the
orientations of the planetary orbits and the stellar rotation axis are
correlated.

\end{abstract}


\section{Introduction}
\label{sec:intro}

The obliquities of the host stars in exoplanetary systems display a
surprising diversity, including low obliquities reminiscent of the
solar system, strongly tilted stars, and retrograde systems in which
the directions of stellar rotation and planetary orbital revolution
are opposite. Most of these results have been obtained by detecting
the Rossiter-McLaughlin (RM) effect, a spectroscopic anomaly that is
observed during a planetary transit (Queloz et al.\ 2000, Winn et
al.\ 2005). In addition, for some systems the obliquity has been
determined through the detection and interpretation of transits of a
planet over starspots (e.g., Deming et al.\ 2011, D{\'e}sert et
al.\ 2011, Nutzman et al.\ 2011, Sanchis-Ojeda et al. 2011), and for
one system (Barnes et al.\ 2011, Szab\'o et al.\ 2011) it has been
estimated using the signatures of gravity darkening from rapid stellar
rotation (Barnes 2009).

Almost all of the previous results pertain to host stars with hot
Jupiters.  The diversity of obliquities seen in those systems has been
taken as evidence that the process of converting a ``normal'' Jupiter
into a hot Jupiter can tilt the inclination of the planetary orbit
(see, e.g., Winn et al.\ 2010a, Triaud et al.\ 2010, Albrecht et
al.\ 2012). It would be interesting to extend these measurements to
systems with longer-period planets, and multiple-planet systems, to
test whether the high obliquities are indeed confined to the hot
Jupiter systems. Unfortunately, the long-period and multiple-planet
systems tend to involve smaller planets and intrinsically fainter host
stars (Latham et al.\ 2011, Steffen et al.\ 2012), making it difficult
to apply the RM and starspot techniques. This is why only two such
systems have been examined to date (Sanchis-Ojeda et al.\ 2012, Hirano
et al.\ 2012b). It would be advantageous to develop a technique that
does not depend so critically on the signal-to-noise ratio of the
transit data.  One possibility is to use a combination of the measured
rotation period ($P_{\rm rot}$), the projected rotation rate ($v\sin
i_{\rm s}$) and the stellar radius ($R$) to determine $\sin i_{\rm
  s}$, the sine of the angle $i_{\rm s}$ between the stellar rotation
axis and the line of sight (see, e.g., Hirano et al.\ 2012a). However,
this method is usually limited by the relatively poor accuracy of
$v\sin i_{\rm s}$ measurements for cool stars.

Asteroseismology provides another potentially powerful method.  The
detection and interpretation of the solar-like oscillations shown by
solar-type stars is well known to provide accurate fundamental
properties of host stars (e.g., Bazot et al.\ 2005, Bouchy et
al.\ 2005, Vauclair et al.\ 2008, Soriano \& Vauclair 2010,
Christensen-Dalsgaard et al.\ 2010, Ballot et al.\ 2011, Batalha et
al.\ 2011, Gilliland et al.\ 2011, Howell et al.\ 2011, Borucki et
al.\ 2012, Carter et al.\ 2012, Escobar et al.\ 2012, Barclay et
al.\ 2013).  Less well known is that in some cases the
rotationally-induced splittings of oscillation modes can be used to
determine $i_{\rm s}$ (Gizon \& Solanki 2003).  When the host star
also has a transiting planet, the inclination $i_p$ of the planetary
orbit can be determined, and therefore the difference in inclination
between the star and planetary orbit can be calculated.

In contrast to the RM and starspot techniques, the applicability of
the asteroseismic method depends predominantly on the stellar
parameters and hardly at all on the planetary parameters, giving the
asteroseismic method a decisive advantage in measuring the stellar
obliquities in systems with small planets or long-period planets. The
asteroseismic analysis does, however, require bright targets and
long-duration, high-cadence photometric time series to give the
requisite signal-to-noise and frequency resolution for extracting
clear signatures of rotation from the oscillation spectrum, and hence
the stellar inclination angle.

Here we present the first application of asteroseismology to the
problem of stellar obliquity determination for Sun-like exoplanet
hosts with transiting planets.  Both of the systems considered in this
paper have solar-type stars hosting multiple, small (super-Earth
sized) transiting planets. The identification of the two-planet
Kepler-50 system (KOI-262, KIC\,11807274), along with the validation
of the transit signals as arising from planets, was previously
reported by Steffen et al.\ (2013). Kepler-65 (KOI-85, KIC\,5866724)
is a three-planet system that is herein identified and validated for
the first time.  Both systems involve F-type stars at the brighter end
of the \emph{Kepler} target list, having apparent magnitudes of
$K_p=10.42$ and $K_p=11.02$, respectively.

Previously, asteroseismic methods have been applied to host stars with
single, non-transiting large planets discovered using the Doppler
method -- HD\,52265, a solar-type host with asteroseismic data from
CoRoT (Ballot et al.\ 2011; and HR\,8799, an A-type host showing
$\gamma$\,Doradus pulsations in ground-based observations (Wright et
al. 2011) -- with only moderate constraints returned on the stellar
inclinations.

The rest of the paper is organized as follows. We begin in
Section~\ref{sec:prop} by estimating the fundamental stellar
properties, using the solar-like oscillations detected in the
\emph{Kepler} lightcurves and complementary spectroscopic data.
Section~\ref{sec:planets} presents the planet properties of both
systems, including validation of the planets orbiting Kepler-65 and
discussion of the mutual inclinations of the planetary orbits of both
systems. The asteroseismic estimation of the stellar obliquities,
which depends on extracting signatures of rotation from the
oscillation spectra, is presented in
Section~\ref{sec:inc}. Section~\ref{sec:surfrot} compares the
asteroseismic results on rotation with independent estimates of the
surface rotation based on the quasi-periodic variations seen in the
\emph{Kepler} lightcurves, and measurements of the sky-projected
surface rotational velocity based on spectroscopic line broadening. We
finish in Section~\ref{sec:disc} with a discussion of the implications
of our results for theories of planetary formation.
 
\section{Fundamental properties of the stars}
\label{sec:prop}

We determined the fundamental stellar properties of Kepler-50 and
Kepler-65 by comparing a few key asteroseismic and spectroscopic
observables to the outputs of stellar-evolutionary models.

The asteroseismic results are based on the \emph{Kepler} short-cadence
(SC) data (Gilliland et al.\ 2010), whose one-minute sampling is
needed to detect the short-period oscillations observed in solar-type
stars (see also Chaplin et al. 2011a).  The lightcurve for Kepler-50
spans 18\,months, from \emph{Kepler} observing quarters 6 through 11
inclusive. The lightcurve for Kepler-65 spans 27\,months, from
quarters 3 through 11.

\begin{figure*}
\epsscale{0.8}
\plotone{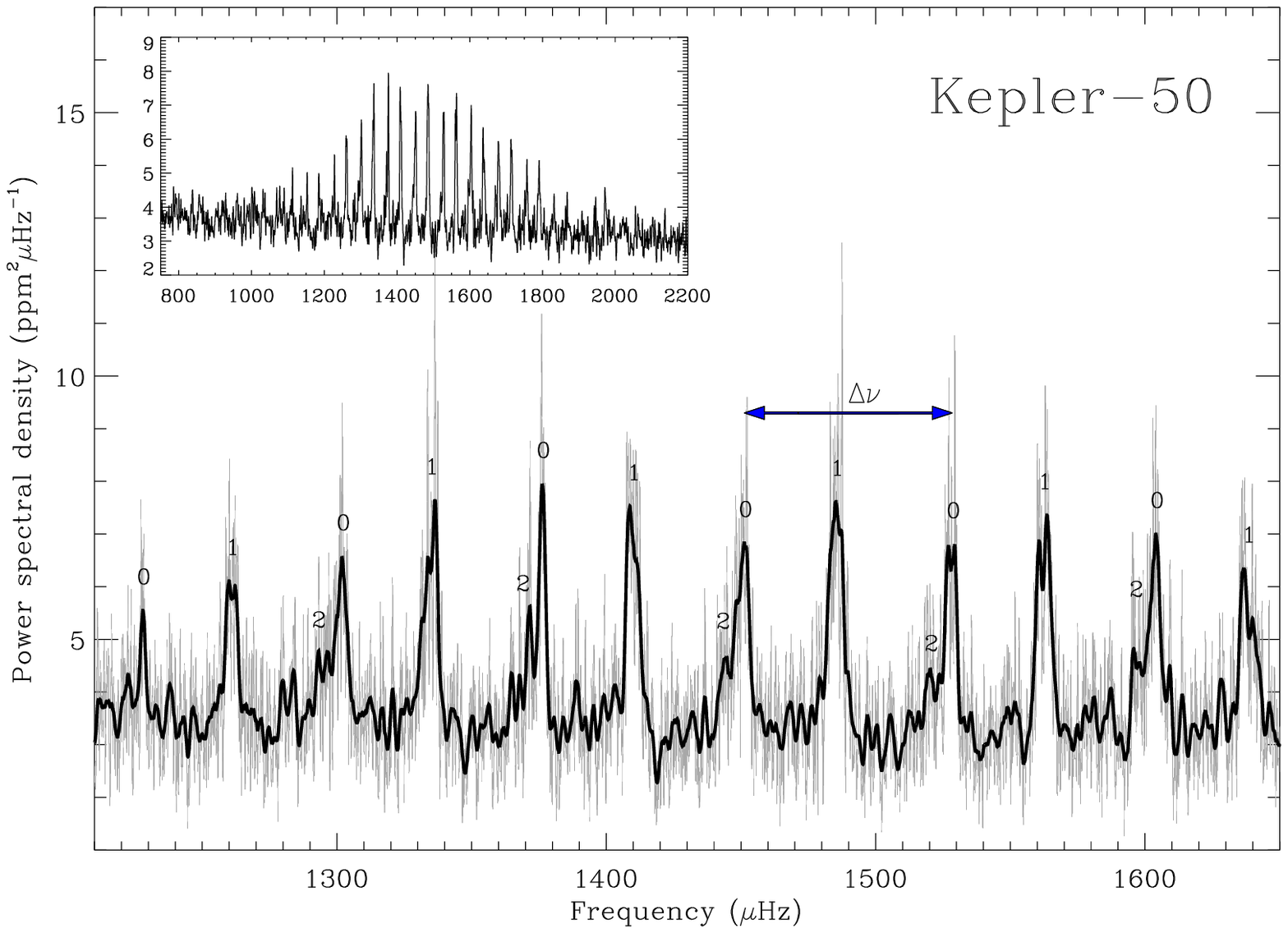}
\epsscale{0.8}
\plotone{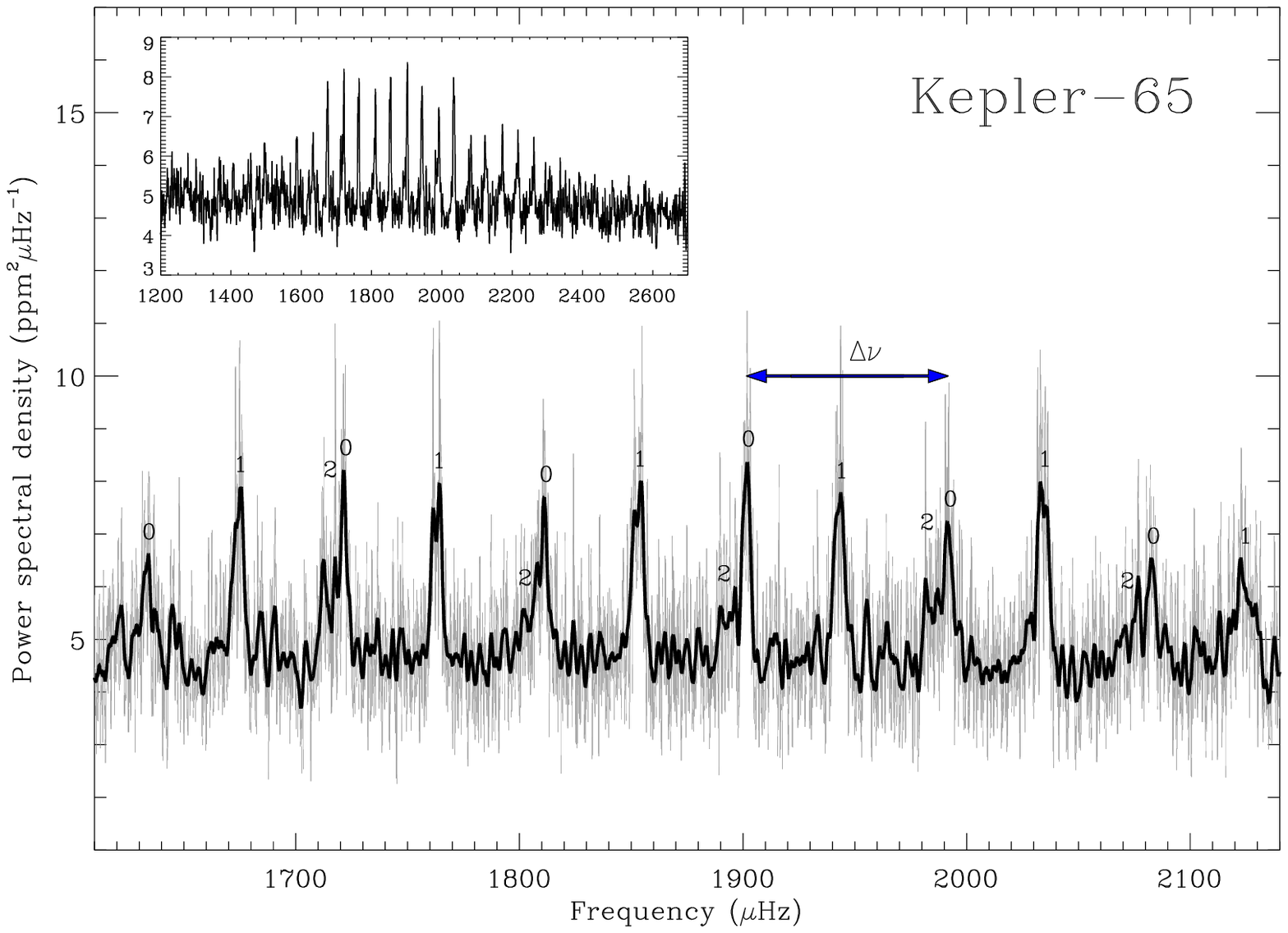}

\caption{Frequency-power spectra of Kepler-50 (top panel) and
  Kepler-65 (bottom panel), showing rich spectra of overtones of
  solar-like oscillations. The main plots in both figures show six
  overtones, with modes tagged according to their angular degree,
  $l$. The so-called large frequency separation between one pair of
  adjacent $l=0$ modes is also marked. The insets show the full
  frequency extent of both observable p-mode spectra. The
  Gaussian-like power envelope of each spectrum is readily apparent,
  which peaks at $\nu_{\rm max}$. Plots rendered in black are the
  power spectra after smoothing with a $1.5\rm \mu Hz$ filter. The
  light grey curves show the spectra after applying lighter
  smoothing.}

\label{fig:powspec}
\end{figure*}

Before computing power spectra, the planetary transit signals were
removed from the time series by applying a median high-pass filter of
width appropriate for the transit durations (see, e.g.,
Christensen-Dalsgaard et al. 2010). The clear separation of the
relevant timescales -- i.e., periods of days associated with the
transits versus periods of minutes associated with the dominant
oscillations -- means that this approach cleans the frequency-power
spectrum in such a way as to allow the asteroseismic analysis to
proceed unhindered.  Fig.~\ref{fig:powspec} shows frequency-power
spectra of the lightcurves of Kepler-50 (top panel) and Kepler-65
(bottom panel). The spectra were computed using a Lomb-Scargle
periodogram (Scargle 1982), and calibrated to satisfy Parseval's
theorem.  Both stars present clear patterns of peaks due to solar-like
oscillations, which are small-amplitude pulsations that are
stochastically excited and intrinsically damped by the near-surface
convection. Many acoustic (pressure, or p) modes of high radial order,
$n$, are excited to observable amplitudes. Solar-type stars oscillate
in both radial and non-radial modes.  The modes may be decomposed onto
spherical harmonic functions of degree $l$. Both stars show detectable
overtones of modes with $l \le 2$.


  \subsection{Spectroscopic data and analysis}
  \label{sec:spectro}

Estimates of $T_{\rm eff}$ and [Fe/H] were obtained by analyzing
high-resolution optical spectra. The observations were made as part of
the \emph{Kepler} Follow-up Observing Program (KFOP). Spectra were
collected for both stars using the HIRES spectograph on the 10-m Keck
telescope on Mauna Kea. In the case of Kepler-50 spectra were also
collected with the fiber-fed Tillinghast Reflector Echelle
Spectrograph (TRES) on the 1.5-m Tillinghast Reflector at the Fred
Lawrence Whipple Observatory, and the Tull Coud\'e Spectrograph on the
2.7-m Harlan J. Smith Telescope at the McDonald Observatory,
Texas. For Kepler-65, additional spectra were collected by the
FIber-fed Echelle Spectrograph (FIES) on the 2.5-m Nordic Optical
Telescope (NOT) on La Palma.

The Keck data were analyzed using the Spectroscopy Made Easy (SME)
pipeline (Valenti \& Piskunov 1996; Valenti \& Fischer 2005). Data
from the other telescopes were analyzed with the Stellar Parameter
Classification (SPC) pipeline (Buchhave et al.\ 2012). Good agreement
was found between the SME and SPC estimates of $T_{\rm eff}$ and
[Fe/H]. For subsequent analysis we adopted the SME values. The SME and
SPC analyses also provided estimates of $v\sin\,i$ based on the
observed line broadening (see Torres et al.\ 2012 for further
details). Section~\ref{sec:surfrot} discusses the comparison of those
results with the asteroseismic estimates of stellar rotation rates.

A well-known problem with the analysis of high-resolution spectra of
solar-type stars is that $\log g$ is difficult to pin down, and
subject to systematic errors that propagate into the uncertainties of
other parameters such as $T_{\rm eff}$ and [Fe/H]. For this reason, an
iterative procedure was used to refine the estimates of the
spectroscopic parameters (e.g., see Bruntt et al.\ 2012, Torres et
al.\ 2012). In this procedure, the initial values of the spectroscopic
parameters are used together with the asteroseismic parameters to
compute log\,$g$ (see next section). The spectroscopic analysis was
then repeated with log\,$g$ fixed at this asteroseismic value, to
yield the revised values of $T_{\rm eff}$ and [Fe/H]. Convergence of
the inferred properties (to within the estimated uncertainties) was
achieved after just a single iteration.  The final, iterated
spectroscopic results are presented in Table~\ref{tab:res}.

  \subsection{Asteroseismic estimation of stellar properties}
  \label{sec:asteroprop}

A two-stage procedure was adopted to estimate the fundamental
properties of the stars, using as input asteroseismic parameters and
complementary spectroscopic results. At the first stage for each star
we sought initial estimates of the stellar properties by searching
among grids of stellar evolutionary models to get a best fit to two
global oscillation properties, the spectroscopically estimated
effective temperature $T_{\rm eff}$, and metallicity, [Fe/H]. The two
asteroseismic properties were $\left< \Delta\nu \right>$, the average
of the large frequency separations between consecutive overtones $n$
of the same angular degree $l$; and $\nu_{\rm max}$, the frequency of
maximum oscillation power. The average large separations scale to very
good approximation as $\left< \rho \right>^{1/2}$, where $\left< \rho
\right> \propto M/R^3$ is the mean density of the star with mass $M$
and surface radius $R$ (see, e.g., Christensen-Dalsgaard 1993). The
frequency of maximum oscillation power has been shown to scale to good
approximation as $gT_{\rm eff}^{-1/2}$ (Brown et al.\ 1991; Kjeldsen
\& Bedding 1995; Belkacem et al.\ 2011), where $g$ is the surface
gravity. Several analysis codes (Christensen-Dalsgaard et al.\ 2010;
Hekker et al.\ 2010; Huber et al.\ 2009; Verner et al.\ 2011) were
applied to the frequency-power spectra to extract the required
estimates.  A final value of each parameter was selected by taking the
individual estimate that lay closest to the median. The uncertainty on
the final value was given by adding (in quadrature) the uncertainty on
the chosen estimate and the standard deviation over the set of
results.  For Kepler-50 we obtained $\left< \Delta\nu \right> = 76.0
\pm 0.9\,\rm \mu Hz$ and $\nu_{\rm max}=1496 \pm 56\,\rm \mu Hz$;
while for Kepler-65 we obtained $\left< \Delta\nu \right> = 90.0 \pm
0.5\,\rm \mu Hz$ and $\nu_{\rm max}=1880 \pm 60\,\rm \mu Hz$.

The grid-based search codes that we then applied to these results are
described by Stello et al.\ (2009), Basu et al.\ (2010), Quirion et
al.\ (2010) and Gai et al.\ (2011). 

In the second stage we used estimates of the individual oscillation
frequencies, along with the revised spectroscopic data, as inputs to a
detailed modelling performed by three members of the team (SB, JCD and
TM). The procedure used to estimate the frequencies -- which also
provided information on the internal rotation and angle of inclination
of each star -- is discussed in detail in Section~\ref{sec:inc}.  More
details on the detailed modeling used to estimate the stellar
properties is given in the Appendix, which followed the methodology
applied in, for example, Christensen-Dalsgaard et al.\ (2010), Howell
et al.\ (2011) and Carter et al.\ (2012). Estimated properties from
the first, grid-based stage were used either as starting guesses or as
a guideline check for initial results.  The final properties presented
in Table~\ref{tab:res} come from the analysis made by JCD (which
provided the median solutions). Uncertainties on the final properties
include a contribution from the scatter between the three different
sets of results. We note that the properties from the first stage
showed excellent agreement with the final estimated properties (i.e.,
to within the estimated uncertainties).


\begin{deluxetable}{cccccccc}
\tabletypesize{\scriptsize} \tablecaption{Estimated stellar properties} 
\tablewidth{0pt}
\tablehead{
 \colhead{Star}&
 \colhead{$T_{\rm eff}$}& \colhead{[Fe/H]}& 
 \colhead{$M$}& \colhead{$R$}& \colhead{$\left< \rho \right>$}&
 \colhead{$\log\,g$}& \colhead{Age}\\ 
 \colhead{}&
 \colhead{(K)}& \colhead{(dex)}&
 \colhead{($\rm M_{\odot}$)}& \colhead{($\rm R_{\odot}$)}& \colhead{($\rm g\,cm^{-3}$)}&
 \colhead{(dex)}& \colhead{(Gyr)}}
 \startdata
Kepler-50& $6225 \pm 66$& $0.03 \pm 0.06$&
                  $1.24 \pm 0.05$& $1.58 \pm 0.02$& $0.441 \pm 0.004$& $4.132 \pm 0.005$& 
                  $3.8 \pm 0.8$\\
Kepler-65& $6211 \pm 66$& $0.17 \pm 0.06$&
                  $1.25 \pm 0.06$& $1.41 \pm 0.03$& $0.621 \pm 0.011$& $4.232 \pm 0.006$& 
                  $2.9 \pm 0.7$
\enddata
\label{tab:res}
\end{deluxetable}


\section{Characterization of the planetary systems}
\label{sec:planets}

It is important to establish whether the transit-like photometric
signals represent actual transits of a system of planets across the
disk of the intended target star, as opposed to a ``false positive''
such as a system of eclipsing stars blended with the intended target
star. For Kepler-50, transit timing variations (TTV) have been
observed for both of its planets and are anti-correlated, a clear sign
that the planets are interacting with each other and hence orbit the
same star (Steffen et al.\ 2013). Kepler-65 has not been confirmed in
this manner; in the following section we validate the system by other
means.

\subsection{Validation of Kepler-65}
\label{sec:validate}

To validate the Kepler-65 system, in this section we will demonstrate
that: (i) background eclipsing binaries are unlikely to be responsible
for any of the three candidate transit signals; (ii) all three
transiting objects are likely orbiting the same star, which must have
a mean density very similar to that of the intended target star; and
(iii) planets c and d are near a 7:5 mean-motion commensurability, and
the smaller planet in this pair (planet d) exhibits a significant TTV
signal of the nature expected for such a configuration.

\subsubsection{Excluding background binary scenarios}

Lissauer et al.~(2012) considered the question of how many of
\emph{Kepler}'s multiple-planet candidates actually represent true
multiple-planet systems, as opposed to unresolved blends of systems
each having only one eclipsing object. For example, a candidate
two-planet system could actually be a single-planet system along
nearly the same line of sight to a background eclipsing binary, or
there could be two eclipsing binaries along the same line of sight
whose eclipses are diluted to planet-like proportions by the constant
light of a foreground star.  Lissauer et al.~(2012) recognized that
false positives of this nature would be randomly distributed among the
target stars, and that the number of multiple-planet candidates is
much larger than would be expected if the candidates were assigned
randomly to target stars.  From this analysis they concluded that the
vast majority of \emph{Kepler}'s multiple-planet candidates do not
represent superpositions of singly-eclipsing systems. For the
population of three-transit candidates such as Kepler-65, Lissauer et
al.~(2012) estimated the chance that at least one of the candidates
represents an unrelated eclipsing system is 0.07\% (an expectation of
0.13 such false positives out of 178 candidates).

These general considerations show that Kepler-65 is very likely a true
multiple-planet system, as opposed to unrelated singly-eclipsing
systems that are blended together in the \emph{Kepler} photometric
aperture. In the remainder of this section we examine the specific
circumstances and follow-up observations of Kepler-65 that also
support this conclusion.

The photometric aperture used for the star changes from quarter to
quarter, but in all cases has a size of approximately $4\times4$
pixels. With a detector scale of 3.98 arcsec pixel$^{-1}$, stars
within a radius of about 12\,arcsec from Kepler-65 could contribute
light to the aperture and could in principle be the source of some of
the transiting signals. We checked for possible contaminating stars
using two different datasets.

Firstly, we consulted the catalog by Adams et al.\ (2012) of adaptive
optics (AO) images of a large sample of KOIs. The range of star
magnitudes that can be detected depends on the distance to the star,
such that one loses the ability to detect faint stars very close to
the main star. These images have a range of 6\,arcsec, outside of
which no information was provided\footnote{The FWHM of the ARIES
  observation used was 0.1\,arcsec.} Only one other star was detected
on the AO image, at a separation of 2.9\,arcsec from Kepler-65. Adams
et al.\ (2012) estimated that the \emph{Kepler} apparent magnitude of
this star is $K_p\approx 21$, i.e., about 10 magnitudes fainter than
Kepler-65.

Secondly, to seek companions outside the 6\,arcsec radius, we
consulted the Naval Observatory Merged Astrometric Dataset (Zacharias
et al.\ 2004). In this catalog, 11 stars are detected within a box of
30\,arcsec centered on the position of Kepler-65. Only three stars
were found that could be candidates for a background blend; one at a
separation of 7.2\,arcsec with an $R$-band magnitude of 18.6 (compared
to 10.5 for Kepler-65), a second star at a separation of 11.5\,arcsec
with an R-band magnitude of 14.1, and a third star at a separation of
11.8\,arcsec with a $B$-band magnitude of 19.3 (11.6 for Kepler-65).

 
\begin{figure*}
\epsscale{1.0}
\plotone{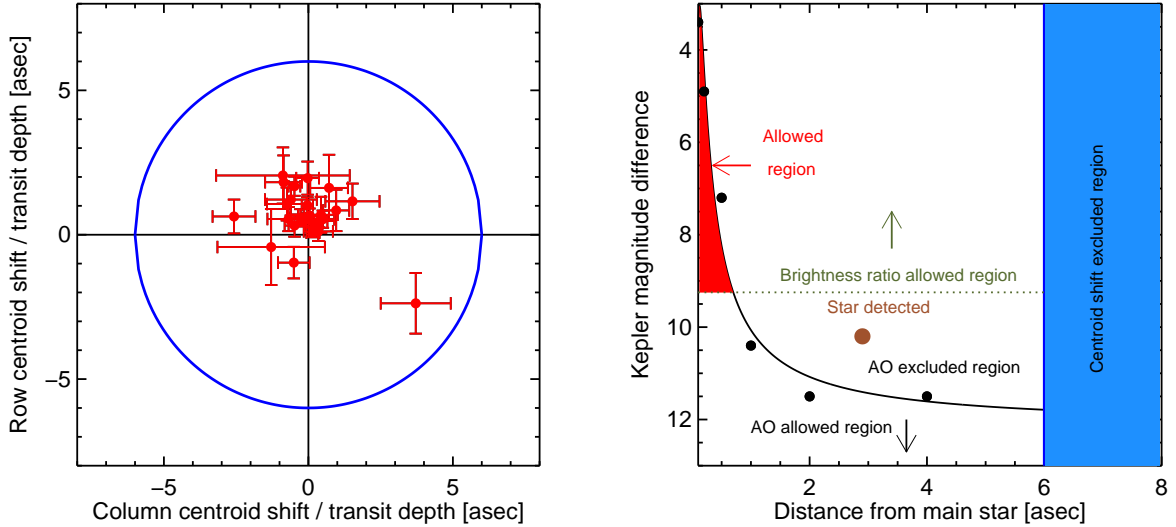}

\caption{Left-hand panel: Centroid shifts during transits divided by
  the transit depth for each candidate during each SC quarter. This is
  an estimate of the distance between the source of the transit signal
  and the center of light of the system. Based on these data the
  transit signal must originate from within a 6\,arcsec radius (blue
  circle; see text). Right-hand panel: parameter space for the
  possible blend scenarios. Plotted on the abscissa is the distance
  from the center of light in the aperture, and on the ordinate the
  difference in magnitude with respect to Kepler-65. In addition to
  the centroid-shift-excluded region (blue), any star just over
  9\,magnitudes fainter than Kepler-65 is excluded because it would
  not contribute enough light on the aperture to produce the observed
  transit depths. This limit is marked by the horizontal dotted line,
  which was computed assuming an eclipse depth of 50\,\%.  The AO
  imaging excludes the region above the continuous black line, which
  was obtained by extrapolation of a best-fitting hyperbolic function,
  fitted to the limits (black dots) given in Table 2 of Adams et
  al. (2012). Only the red region is still allowed; in this sense the
  ``radius of confusion'' is 0.7\,arcsec.}

\label{fig:validation}
\end{figure*}


If one of these objects were a background binary star mimicking a
transiting planet, then the spatial, first-moment centroid of the
light gathered on the aperture would be displaced during eclipses by
an amount approximately equal to the observed transit depth multiplied
by the projected distance from the object to Kepler-65. Therefore,
since the transit depth is known independently with high precision, an
upper bound on the centroid displacement can be used to set a maximum
distance at which a contaminating binary can be located (also known as
the radius of confusion). This notion has been applied to detect
background binaries among the KOIs (see, e.g., Batalha et al.\ 2010),
to estimate false alarm probabilities (FAPs) for particular KOIs
(Morton \& Johnson 2011) and to validate individual candidates using
the BLENDER technique (Torres et al.\ 2011).

In the case of Kepler-65, one pixel of the stellar image is saturated,
and consequently the distribution of light does not follow the
standard point-spread function. Rather than attempting to model the
saturated point-spread function, we used the flux-weighted column and
row centroids produced by the \emph{Kepler} pipeline. With this method
of computing centroids we are only sensitive to displacements larger
than $\approx$1~pixel ($\approx$4\,arcsec), but this is sufficient for
our purpose. For each transit observed at short cadence (SC), we
selected a window in time of width 4.8\,hr centered on each
transit. Transits that occurred within 6\,hr of another were
excluded. To eliminate the effects of outliers we omitted data points
differing by more than 3$\sigma$ from a median-smoothed version of the
time series, where the smoothing was performed over 30-min
intervals. We found that the centroid motion was approximately a
linear function of time, presumably because of the continuous pointing
drift of the telescope. We corrected for this effect by fitting the
out-of-transit portions of the dataset with a linear function of
time. All the centroid information for each candidate in a given
quarter was then phase-folded using a linear transit ephemeris from
the KOI input catalog (Batalha et al.\ 2012). The sequences of
in-transit and out-of-transit centroids were approximated as Gaussian
distributions, and the centroid displacement was computed as the
difference between the means of the distributions, with an uncertainty
based on standard error propagation.

The left-hand panel of Fig.~\ref{fig:validation} shows the measured
centroid displacements after dividing by the corresponding transit
depths, so that implied physical distances are plotted. The signals
cannot originate from a source outside the 6\,arcsec radius (shown in
blue) since no points lie outside that range. The right-hand panel of
Fig.~\ref{fig:validation} shows this constraint, in combination with
constraints from other considerations, which together limit the radius
of confusion to 0.7\,arcsec. One constraint is the AO imaging
described previously. Another is that binary stars just over
9\,magnitudes fainter than Kepler-65 cannot decrease the total amount
of light by 100\,ppm (the depth of the shallowest transit), even were
they to have eclipses of 50\,\% depth, i.e., as given by $\Delta K_p =
-2.5 \log_{10} \left[100 \times 10^{-6} \right / 0.5] \simeq
9.2$. This limit is marked as the horizontal dotted line in right-hand
panel of Fig.~\ref{fig:validation}. No stars brighter than that limit
and within 0.7 to 6\,arcsec from Kepler-65 were detected within the AO
image. With such a small radius of confusion, the probability that any
of the signals come from background binaries is small. We estimate
this probability to be $\leq$0.15\,\% for each candidate of Kepler-65,
based on the work by Morton \& Johnson (2011), who calculated the
local surface density of eclipsing binaries whose properties could
mimic those of each \emph{Kepler} candidate. Specifically we took
their estimated false-positive probability of $\leq$1\,\%, which
assumed a radius of confusion of 2\,arcsec, and scaled it by
$(0.7/2.0)^2$ (because we have demonstrated that the true radius of
confusion for Kepler-65 is 0.7\,arcsec).

Finally there is the ``multiplicity boost,'' as discussed at the
beginning of this section.  Since there are three transit candidates
for a single \emph{Kepler} target, the false alarm probability of each
individual transiting object is further reduced, according to the
statistical argument of Lissauer et al.\ (2012). Here the boost factor
is approximately 50, which would reduce the individual false-positive
probabilities from $\leq$0.15\% to $\leq$$3\times 10^{-5}$.

\subsubsection{Evidence that the three planets orbit Kepler-65}
\label{subsec:evidence-3planets}

We have demonstrated that it is unlikely that any of the transit
candidates arises from a background eclipsing binary. We next ask
whether the signals represent three planets all orbiting the intended
target star Kepler-65, or whether any of them could actually be
orbiting a companion star that is gravitationally bound to it. To
address this question we searched for a pattern in the transit
observables that would suggest that all the planets orbit the same
star; namely, transit durations scaling as the cube root of the
orbital period, which is a sign that they transit a star with similar
density (see, e.g., Fabrycky et al.~2012, Lissauer et al.~2012). If
some of the transit signals represented transits across a different
star, then such a pattern would occur only by coincidence.

To measure transit durations, we constructed phase-folded transit
light curves for each of the three candidates using the SC data and
assuming a constant orbital period. Transits that occurred within
6\,hr of another were excluded, removing all possible overlapping
transits. The data were binned into 7.5-sec intervals to increase the
speed of subsequent computations. We used a standard description of
the loss of light due to a transiting planet (Mandel \& Agol 2002) to
model the binned light curves simultaneously. We adopted a quadratic
limb-darkening law, with the two coefficients left as free parameters
(and shared by all three candidates). The free parameters describing
each light curve were the squared planet-to-star radius ratio $(R_{\rm
  p}/R)^2$, the impact parameter, $b$, and the stellar radius divided
by the orbital distance, $R/a$. We then found the best-fitting model
parameters that minimized the standard $\chi^2$ function, with
uncertainties on the measurements defined as the standard deviation of
the points outside transit for each of the folded light curves. A
Markov Chain Monte Carlo (MCMC) code was then used to explore the
range of allowed parameters.

\begin{figure*}
 \epsscale{1.0}
 \plotone{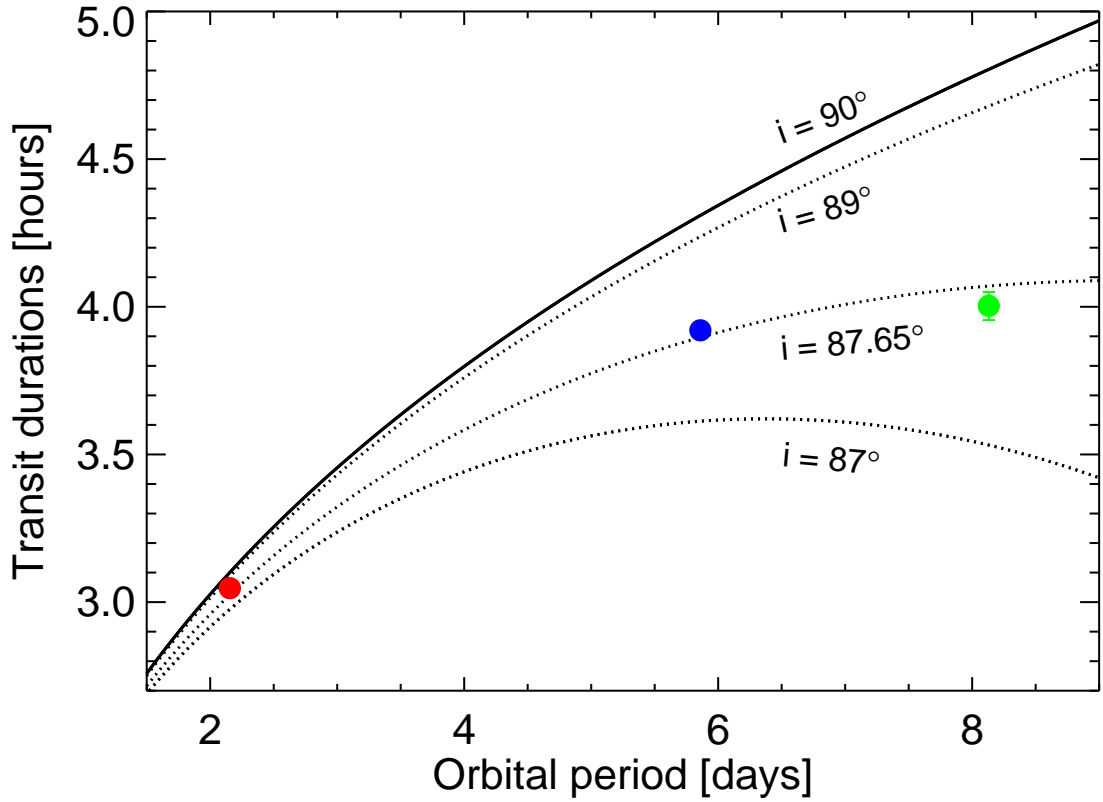}

 \caption{Measured transit durations of the three planets orbiting
   Kepler-65 (filled colored circles). The solid black line shows the
   expected durations for planets transiting Kepler-65 in circular
   orbits with inclination 90$^\circ$ (zero impact parameter). The
   dashed lines show the durations for different orbital
   inclinations. The durations are consistent with the three planets
   orbiting Kepler-65 in coplanar circular orbits.}

\label{fig:inclination}
\end{figure*}

From the best-fitting model parameters we computed the transit
durations, defined as the interval over which the center of the planet
is projected in front of the stellar disk. This parameter is generally
well constrained, and does not change much in the presence of a small
TTV signal (whereas the ingress duration would experience larger
fractional variations). The transit durations are plotted in
Fig.~\ref{fig:inclination}, as a function of orbital period. We
compared these values with those expected for planets in circular
orbits around a star with a mean density equal to
$0.621$~g\,cm$^{-3}$, which is the mean density of Kepler-65 as
estimated from the asteroseismic analysis (see Table~1 and Section~2).
The measured durations agree well with a model in which all planets
have the same orbital inclination, which is good evidence that the
planets have nearly coplanar and circular orbits around a single star
with a density similar to that of Kepler-65.


Another way to perform this test is to use the transit observables to
compute the implied mean density of the host star, and compare the
result to the mean density obtained from asteroseismology. To this end
we performed a second fit to the data in which the $R/a$ value for
Kepler-65c (the candidate with the highest signal-to-noise ratio) was
a free parameter, and the $R/a$ values for the other two planets were
fixed according to the assumption of circular orbits around the same
star (i.e., by scaling according to orbital period and Kepler's Third
Law). This effectively introduces a constraint that all three
planetary signals agree on the mean stellar density (Seager \&
Mall{\'e}n-Ornelas 2003). We found the photometrically-derived mean
density to be $0.57^{+0.06}_{-0.07} \,\rm g\,cm^{-3}$, in agreement
with the asteroseismically derived mean density of $0.621 \pm
0.011\,\rm g\,cm^{-3}$ (see Table~\ref{tab:res} and
Section~\ref{sec:prop}). Therefore, the transit observables are
consistent with a system of three planets on nearly circular orbits
around a star with the same mean density as the star that is the
source of the observed p-mode oscillations.

A devil's advocate would raise the possibility that this agreement is
a coincidence, and that one or more of the planets actually orbit a
secondary companion star. This seems unlikely indeed although we do
not attempt to assign a quantitative false positive probability to
this scenario. To establish the probability of such a coincidence one
would need to consider a realistic distribution of companions, along
with their planets, transit probabilities (which may be correlated
with the transit probabilities of the primary star), and transit
durations. One would then need to exclude cases in which the companion
would have been detectable in the optical spectrum, the spectrum of
p-mode oscillations\footnote{The asteroseismic analysis allows us to
  rule out the presence of a bound companion having the same density
  as Kepler-65, since we would have detected a second set of
  oscillations in the frequency-power spectrum, overlapping in
  frequency with the oscillations of Kepler-65. In fact, given the
  observed background noise level, and using the asteroseismic
  detection prediction code in Chaplin et al. (2011b), we can rule out
  a bound companion having a density up to $\approx 1.5$-times that of
  Kepler-65 (since it would still have shown detectable
  oscillations).}  (i.e., by contributing signatures of its own
oscillations), or through excessively diluted transit depths, and then
compute the integrated probability of the allowed phase space. This is
beyond the scope of this study.

\subsubsection{Detection of a TTV signal for Kepler-65d}

The detection of transit timing variations have proven to be useful
for validating planets as well as constraining the masses of the
transiting planets. We performed a transit-timing analysis of
Kepler-65 as follows. To measure individual transit times for each
planet we employed a phase-folded light curve as a template function.
Specifically we used a phase-folded light curve that was obtained by
fitting all of the transits under the assumption of a circular orbit
with a constant period. The template was then fitted to the data from
each transit observed at SC, with three free parameters: the central
time of the transit, the out-of-transit flux level, and a constant
gradient in the out-of-transit flux level. An MCMC code was used to
obtain the posterior distribution for the time of transit. Since no SC
data were available in Q0, Q1, or Q2, for those quarters we used the
transit times from the \emph{Kepler} TTV catalog (Rowe et al., in
preparation), measured as described by Ford et al.\ (2011).

The individual transit times were then fitted with a linear function
of epoch, and this function was subtracted from the timing data to
isolate any timing residuals. A visual inspection showed no obvious
TTV signal in the residuals. We fitted these residuals with a
sinusoidal model with three parameters: a TTV period, phase, and
amplitude. To facilitate the exploration of the parameter space, we
divided the range of periods into small intervals covering periods
from 10 to 500\,days, and for each period we optimized the other two
parameters. For the three planets, the best-fitting sinusoids gave
unacceptably high $\chi^2$ values relative to the number of degrees of
freedom (2200, with 356 points for planet b, 198 with 121 points for
planet c, and 200 with 81 points for planet d). The uncertainties on
the individual transit times are likely underestimated due to
correlated noise in the photometric time series (due to some
combination of stellar granulation, p-modes, and instrumental
noise). We proceeded by enlarging the uncertainties by a scale factor
(see Table \ref{tbl:TTV}) such that the minumum $\chi^2$ was equal to
the number of degrees of freedom.

To search for a sinusoidal TTV signal for each of the planets, we used
the Bayesian information criterion. The criterion requires that when
$k$ new parameters are introduced in a model, one needs to achieve a
decrease in $\chi^2$ larger than $k \ln{N}$, where $N$ is the number
of data points, to justify the addition of the extra
parameters. Table~\ref{tbl:TTV} shows that the additional three
parameters are only justified for Kepler-65d, and not for the other
two planets. That planet d is singled out in this test is consistent
with the hypothesis that all three planets orbit Kepler-65, as this
planet has the longest orbital period and is closest to the largest
and presumably most massive planet (Kepler-65c; see
Section~\ref{sec:subtran} and Table~\ref{tbl:Planetary}), factors
which enhance the amplitude of the TTV signal. Its orbital period is
1.39 times that of Kepler-65c, making it near a 7:5 ratio. This has
been observed in many other multiplanet systems (Fabrycky et
al.\ 2012).

Even though the TTV signal of Kepler-65c did not satisfy the Bayesian
information criterion for detection, the measured TTV period of the
best-fitting sinusoid is close to the value expected from the formula
of Agol et al.\ (2005), which for the c and d pair is 50\,days (in
agreement with the detected period). With more data one might be able
to establish this signal more securely. Using the 0.9-min amplitude of
this hypothetical signal as a reference, one would estimate a mass (or
upper bound; see Lithwick et al.\ 2012) for Kepler-65d of
approximately 10 $M_{\rm E}$.

We conclude that Kepler-65 is indeed transited by a system of three
planets, based on the low false-alarm probability for each individual
transit, the unlikely coincidence that would be required for a
spurious system to produce the observed trend of transit durations
versus orbital periods, the agreement between the photometric and
asteroseismic estimates of the mean stellar density, and the detection
of a physically reasonable TTV signal for at least one of the
planets. However, as in many other cases, we acknowledge the fact that
we cannot completely rule out the unlikely companion scenario, in
which one or more of the planets orbits a fainter, bound companion
having a higher density than Kepler-65.


 \begin{deluxetable}{cccccccc}
 \tabletypesize{\scriptsize}
 \tablecaption{TTV signals of Kepler-65\label{tbl:TTV}}
 \tablewidth{0pt}

 \tablehead{
 \colhead{Planet} &
 \colhead{$N$} &
 \colhead{Error scale} &
 \colhead{Period} &
 \colhead{Amplitude} &
 \colhead{$\Delta \chi^2$ } &
 \colhead{$k \ln{N}$}&
 }

 \startdata

b & 356 & 2.5 & 32.6 days &  4.6 min & 13.9 & 17.6 \\
c & 121 & 1.3 & 49.7 days &  0.9 min & 8.8 & 14.4 \\
d & 81 & 1.6 & 44.5 days &  9.2 min & 31.9 & 13.2 \\

 \enddata

 \tablecomments{Summary of the search for sinusoidal TTV signals in
   the transit times of Kepler-65b, c and d. After correcting the
   timing error bars to account for correlated noise and low S/N, we
   find that only Kepler-65d has a significant detection.}

 \end{deluxetable}


\subsection{Transit parameters for Kepler-50 and Kepler-65}
\label{sec:subtran}

Transit timing variations are quite significant for Kepler-50. For
this reason, care was needed in producing a phase-folded transit light
curve for subsequent analysis. In addition to the three transit model
parameters ($R/a$, $b$, $(R_{\rm p}/R)^2$) for each candidate and the
two limb-darkening coefficients, we modelled the interval between
transits as a constant plus a sinusoidal function of time.  We fitted
this model to SC data (Q6 through Q11) for transits separated from
each other by at least 6\,hr and, using the best-fitting model, we
folded the data and binned it to a cadence of 7.5\,sec. This template
was then used to obtain the transit timings with uncertainties as
described in the previous section. The new TTV signal, including the
long-cadence (LC) timings from the \emph{Kepler} catalog, was then
fitted to improve the sinusoidal component of the ephemeris, which in
turn was used to properly fold the data. This iterative process
converged when the sinusoidal component did not change significantly
from one step to the next.

 
\begin{figure*}
\epsscale{1.0}
\plotone{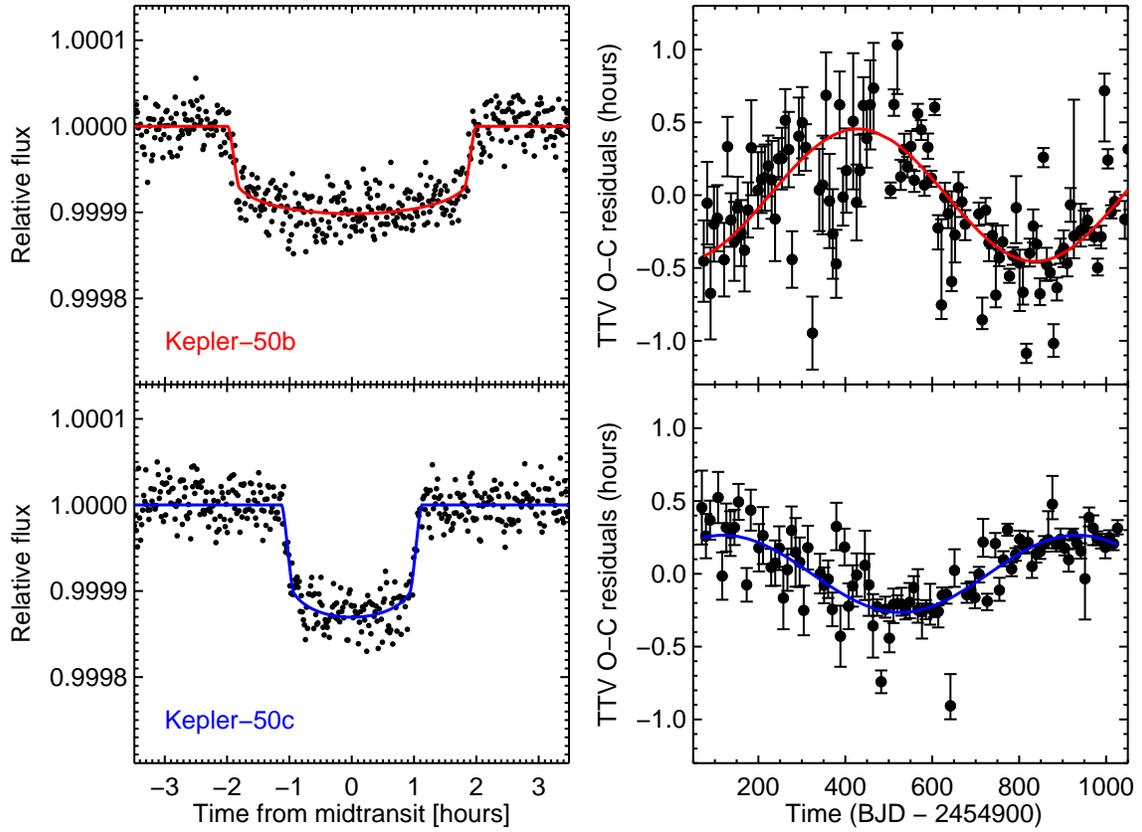}

\caption{Left-hand panels: the black dots show the binned SC
  (one-minute cadence) data for each of the Kepler-50 planets, and the
  lines show the best-fitting transit models. Right-hand panel: TTVs
  and uncertainties. The sinusoidal anti-correlated signals are
  plotted with thick lines.}

\label{fig:Kepler50}
\vspace{0.5cm}

\end{figure*}


The final phase-folded light curves and the best-fitting models are
shown in Fig.~\ref{fig:Kepler50}. The two planets have very similar
orbital periods, with a period ratio close to 1.2. We see in the
figure that the transits of the outer planet are much shorter in
duration than those of the inner planet. This indicates that the
transits of planet c have a high impact parameter. In this situation
there is a risk of bias in the determination of the planet radius due
to poorly constrained limb-darkening coefficients. To avoid this, we
introduced Gaussian priors on each coefficient with values of $0.3 \pm
0.1$, based on the theoretical coefficients given by Claret \& Bloemen
(2011) for stars similar to Kepler-50. Since the orbits are so close
to each other, we assumed circular orbits around the same star,
essentially linking all of the $R/a$ parameters for the two planets
(see section \ref{subsec:evidence-3planets}). The resulting planet
parameters are given in Table~\ref{tbl:Planetary}. The stellar density
derived from this transit model has a large uncertainty due to the low
S/N of the transits, but the final value $0.40^{+0.6}_{-0.10}\,\rm
g\,cm^{-3}$ is nevertheless compatible with the much more precisely
determined asteroseismic density of $0.441 \pm 0.004\,\rm g\,cm^{-3}$
(see Table~\ref{tab:res} and Section~\ref{sec:prop}).


 \begin{figure*}
 \epsscale{1.0}
 \plotone{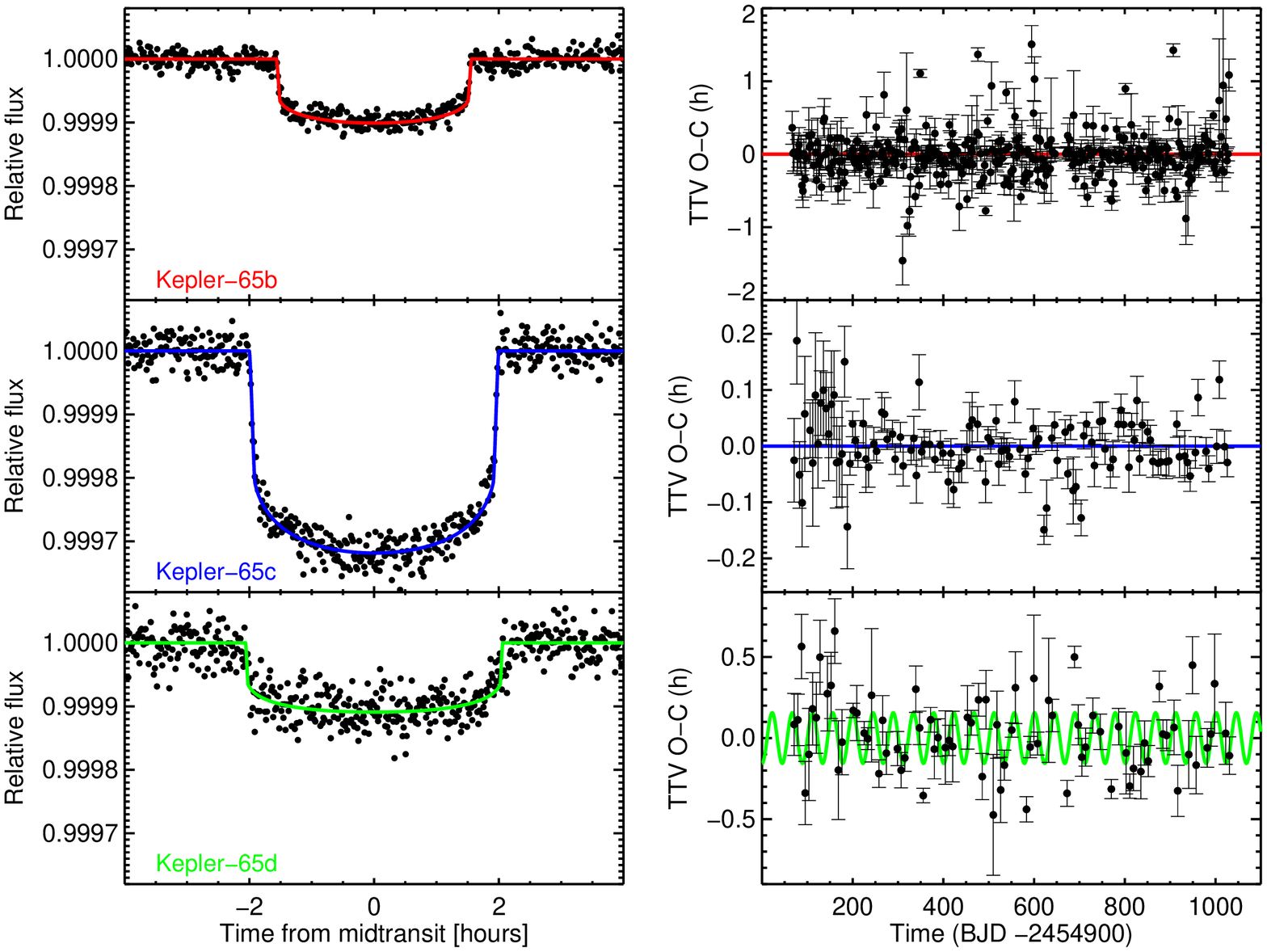}

\caption{Similar to Fig.~\ref{fig:Kepler50} but for the Kepler-65
  system.}

 \label{fig:Kepler65}
 \end{figure*}




 \begin{deluxetable}{l p{0.5cm} cc p{0.5cm} cccccc}

 \tabletypesize{\scriptsize}
 \tablecaption{Transit parameters \label{tbl:Planetary}}
 \tablewidth{0pt}

 \tablehead{
 \colhead{Parameter} &
 \colhead{} &
 \colhead{Kepler-50b } &
 \colhead{Kepler-50c} &
 \colhead{} &
 \colhead{Kepler-65b} &
 \colhead{Kepler-65c} &
 \colhead{Kepler-65d}&
 }

 \startdata

$(R_{\rm p}/R)^2$ [ppm] & &$99^{+5}_{-11}$  & $159^{+10}_{-11}$ & & $85.0^{+1.6}_{-1.1}$ & $282^{+5}_{-2}$    & $98^{+2}_{-2}$  \\ 
Impact parameter b & & $0.74^{+0.07}_{-0.36}$ & $0.94^{+0.02}_{-0.06}$ & & $0.16^{+0.20}_{-0.11}$ & $0.42^{+0.10}_{-0.02}$ & $0.53^{+0.07}_{-0.02}$ \\
$R/a$  & & $0.095^{+0.015}_{-0.025}$ & $0.084^{+0.013}_{-0.022}$ & & $0.188^{+0.011}_{-0.002}$ & $0.097^{+0.006}_{-0.001}$ & $0.078^{+0.005}_{-0.001}$ \\
LD coefficient $u_1$& & $0.28^{+0.08}_{-0.08}$ & -- & & $0.25^{+0.07}_{-0.08}$ & -- & -- \\
LD coefficient $u_2$ & & $0.29^{+0.08}_{-0.07}$ & -- & & $0.37^{+0.11}_{-0.10}$ & -- & -- \\ 
Transit duration $T_{1.5-3.5}$ [hours] & &  $3.80^{+0.05}_{-0.03}$   &  $2.06^{+0.03}_{-0.03}$  &  &  $3.077^{+0.07}_{-0.007}$   &  $3.928^{+0.008}_{-0.007}$  &  $4.10^{+0.02}_{-0.03}$ \\

Orbital period [days] & & $7.81254(10)$ & $9.37647(4)$ & & $2.154910(5)$  & $5.859944(3)$ & $8.13123(2)$\\
Time of transit [BJD-2454900] & & 74.376(7) & 69.958(3) & & 66.4990(13) & 65.0391(3) & 70.9905(16)  \\
Planet radius [$R_E$] & & $1.71^{+0.05}_{-0.10}$ & $2.17^{+0.07}_{-0.08}$ & & $1.42^{+0.03}_{-0.03}$ & $2.58^{+0.06}_{-0.06}$ & $1.52^{+0.04}_{-0.04}$\\
Semi-major axis [AU] & & $0.077^{+0.012}_{-0.020}$ &
$0.087^{+0.014}_{-0.023}$ & &  $0.035^{+0.002}_{-0.001}$ &
$0.068^{+0.004}_{-0.002}$ &  $0.084^{+0.006}_{-0.002}$ 

 \enddata

\tablecomments{Summary of planetary parameters. The first five
  parameters were estimated from the folded light curve analysis, and
  the durations were obtained from those parameters. Uncertainties
  come from an MCMC analysis. Orbital periods and times of transit
  come from a fit to the transit times, with a sinusoidal component in
  the case of Kepler-50 and only a linear term for
  Kepler-65. Estimation of planetary radii and semi-major axes also
  made use of the stellar radii from the asteroseismic analysis.}

 \end{deluxetable}



For Kepler-65 we used the analysis discussed in the previous section
to construct the phase-folded light curves. Since no TTV signal was
detected for planets b and c, a constant period was assumed in
constructing the phase-folded light curves based on SC data. For
planet d, the best-fitting sinusoidal TTV model was used to fold the
transits. An individual analysis for each planet showed that the
ingress duration of planet d was still larger than expected, by a
factor of about two. This long ingress duration implies a large impact
parameter, which again leads to large uncertainty and possible bias in
the planetary radius. As for Kepler-50, we assumed the planets to be
on circular orbits around the same star. The phase-folded light curves
of all three transiting planets are shown in Fig.~\ref{fig:Kepler65},
along with the best-fitting models.

Final values of the planet parameters are presented in
Table~\ref{tbl:Planetary}. With the improved folded light curve for
planet d, the stellar mean density (assuming circular orbits) is
$0.61^{+0.02}_{-0.10}\,\rm g\,cm^{-3}$, which also agrees with the
value obtained from asteroseismology. One could use the asteroseismic
density as a prior on our model, but this would not necessarily lead
to greater accuracy because non-zero eccentricities cannot be ruled out
for this system.

Finally, we note that in the transit analysis for both Kepler-50 and
Kepler-65 we have assumed that the light from blended stars is
negligible (i.e.\ a contamination factor of zero). This is well
justified by the AO images (Adams et al.\ 2012). The contamination
factors given in the \emph{Kepler} Input Catalog are very low for both
systems, and the uncertainties in the planetary radii are dominated by
statistical uncertainties rather than the systematic effects of
possible contamination.

\subsection{Discussion of the coplanarity of the systems}

In order to interpret the measured stellar obliquity in the context of
the formation and evolution of the system, it is important to decide
whether the planets are in coplanar orbits (Sanchis-Ojeda et
al.~2012). The low S/N of the transit signals makes it difficult to
use transit observables to constrain the mutual inclination, but the
fact that we have found several planets transiting each star already
tells us that these systems are likely to be coplanar (Lissauer et
al.\ 2011). The probability that a planet on a randomly oriented
circular orbit will transit a star is given by $R/a$. Using the values
from Table~\ref{tbl:Planetary} we may evaluate the probability for two
extreme cases: (i) the planets have coplanar orbits; and (ii) their
orbital orientations are uncorrelated. In the first case, if the most
distant planet transits the star, the other planets in the system will
also transit and so the probability of all planets transiting is equal
to $p_1 = R/a_q$, where $q$ refers to the most distant planet. If the
planets' orbits have independent random orientations, then the
probability $p_2$ that all planets will transit is then equal to the
product of the individual probabilities for each planet. Evaluating
these probabilities $p_1$ and $p_2$ for Kepler-50, we find that
$p_1/p_2 = 10.5$, i.e., the likelihood for coplanar orbits is ten-times
higher. For Kepler-65 the ratio is 55, favoring coplanar orbits even
more strongly. Thus, both systems are likely to be nearly coplanar,
although moderate mutual inclinations cannot be ruled out by this
analysis. More definitive results might eventually be achieved through
transit-timing studies or the detection of planet-planet eclipses
(see, e.g., Hirano et al.\ 2012b).

\section{Asteroseismic determination of stellar angle of inclination}
\label{sec:inc}

\subsection{Principles of the method}
\label{sec:princ}

Asteroseismic estimation of the stellar angle of inclination, $i_{\rm
  s}$, rests on our ability to resolve and extract signatures of
rotation in the non-radial modes from the oscillation spectrum.
Detailed descriptions of the principles of the asteroseimic method may
be found in Gizon \& Solanki (2003) and Ballot et al. (2006,
2008). Here, we summarize the key points.

Rotation lifts the degeneracy in the oscillation frequencies
$\nu_{nl}$, so that the frequencies of non-radial modes ($l>0$) depend
on the azimuthal order, $m$. For the fairly modest rates of rotation
typical of solar-like oscillators we may ignore, to first order, the
effects of the centrifugal distortion (e.g., see Reese et al.\ 2006;
Ballot 2010).  The $2l+1$ rotationally split frequencies may then be
written:
 \begin{equation}
  \nu_{nlm} \equiv \nu_{nl} + \delta\nu_{nlm},
 \end{equation}
with
 \begin{equation}
 \delta\nu_{nlm} \simeq \frac{m}{2\pi}
                \int_0^R \int_0^\pi K_{nlm}(r,\theta) 
                \Omega(r,\theta)r\,{\rm d}r\,{\rm d}\theta.
 \label{eq:rot}
 \end{equation}
Here, $\Omega(r,\theta)$ is the position-dependent internal angular
velocity (in radius $r$, and co-latitude $\theta$), and $K_{nlm}$ is a
weighting kernel that reflects the sensitivity of the mode to the
internal rotation as a function of depth. For modest rates of
differential rotation (in latitude and radius) and absolute rotation,
the splittings $\delta\nu_{nlm}$ of the observable high-$n$, low-$l$ p
modes will take very similar values, hence tending to the
approximation of solid-body rotation (Ledoux 1951). Here, we found no
evidence for significant mode-to-mode variation of the frequency
splittings in the oscillation spectrum of either star.  In what
follows we therefore modelled all splittings as being equal, i.e.,
$\delta\nu_{nlm} = \delta\nu_{\rm s}$. The above also neglects any
contributions to the splittings from near-surface magnetic fields,
which give rise to frequency asymmetries of the observed
splittings. The levels of activity in both stars -- as revealed by
signatures of rotational modulation of spots and active regions in the
\emph{Kepler} lightcurves -- are notably lower than those displayed by
the active Sun (see later, in Section~\ref{sec:surfrot}). Since
magnetic contributions to the solar low-$l$ splittings are small in
size and very hard to measure in Sun-as-a-star data of much higher S/N
(e.g., see Gough \& Thompson 1990; Chaplin 2011, and references
therein), asymmetries here should not be a cause for concern for the
analysis.


\begin{figure*}
\epsscale{0.5}
\plotone{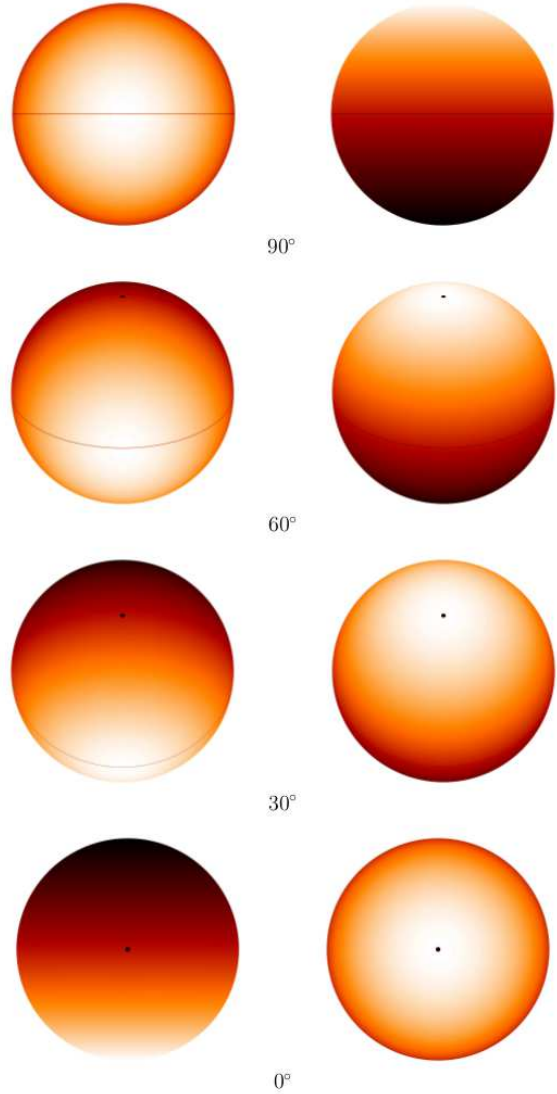}

\caption{Intensity perturbations for $l=1$ mode components, at a phase
  corresponding to extreme displacement of the oscillations. Plotted
  are patterns for $m=1$ (left-hand column) and $m=0$ (right-hand
  column) modes viewed at different angles, $i_{\rm s}=90^\circ$ (top
  row), $60^\circ$ (second row), $30^\circ$ (third row) and $0^\circ$
  (bottom row). The filled circles mark the pole of the rotation axis
  and the lines the stellar equator.}

\label{fig:spher}
\end{figure*}


\begin{figure*}
\epsscale{1.0}
\plotone{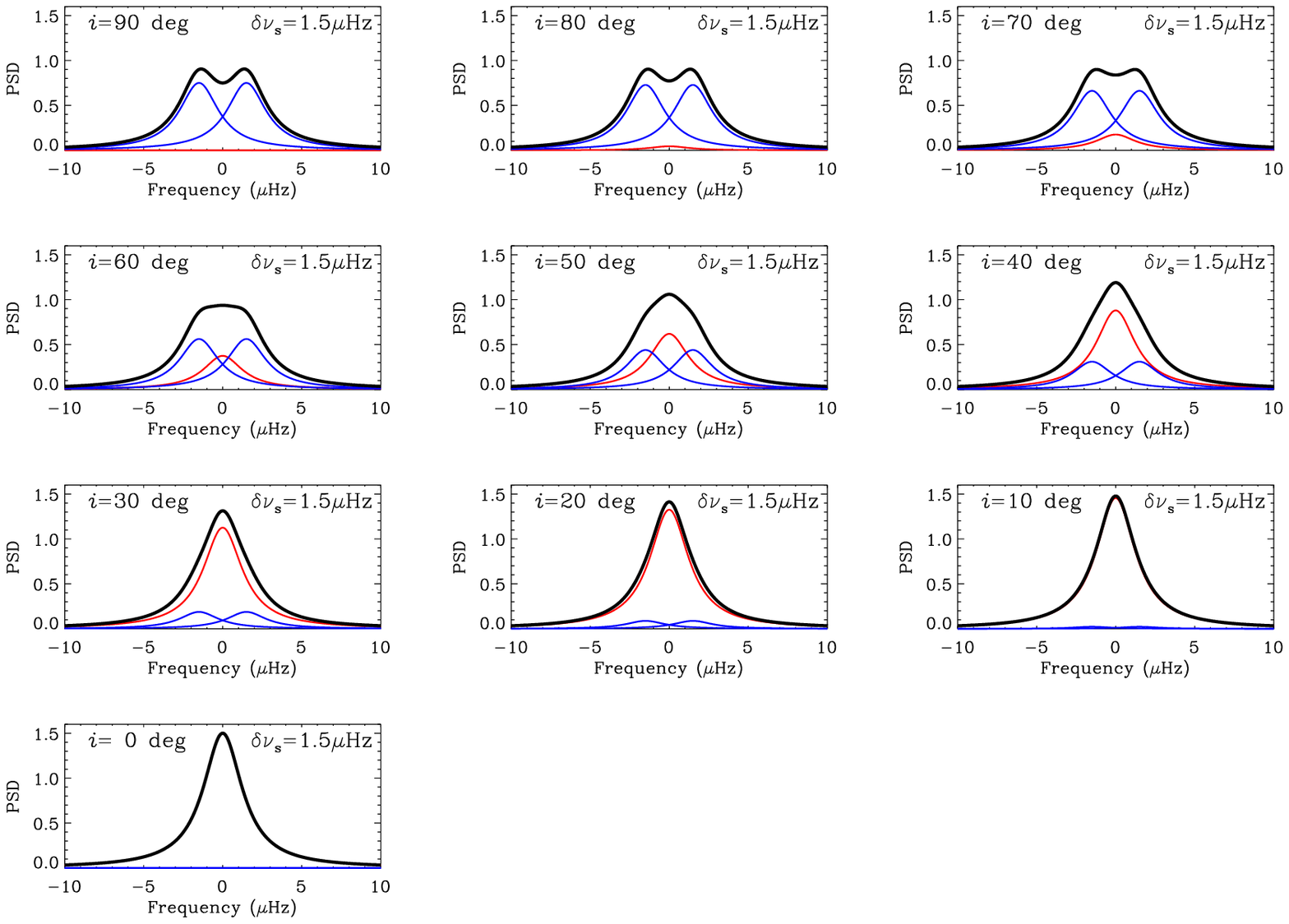}

 \caption{Theoretical profiles of an $l=1$ mode observed at different
   stellar inclination angles, $i_{\rm s}$. The $m= \pm 1$
   components are plotted in blue, the $m=0$ components in
   red, and the thick black line shows the combined multiplet profile.
   The peak linewidth of each component is $\Gamma=3.0\,\rm \mu Hz$
   and the rotational frequency splitting is $\delta\nu_{\rm
     s}=1.5\,\rm \mu Hz$. Panels in the left-hand column show the
   appearances of the multiplet for each of the angles illustrated in
   Fig.~\ref{fig:spher}.}

 \label{fig:example} 
\end{figure*}


The determination of the inclination of the stellar rotation axis
relies on the fact that the mode patterns of the non-radial modes are
not spherically symmetric.  The disk-integrated amplitudes of the $m$
components in any given non-radial muliplet will therefore depend on
the viewing angle.  Fig.~\ref{fig:spher} shows a snapshot of the
intensity perturbations of the $m=1$ (left-hand column) and $m=0$
(right-hand column) components of an $l=1$ mode viewed at different
angles, $i_{\rm s}$.  The perturbations are shown at a phase
corresponding to extreme displacement of each oscillation mode. The
filled circles mark the pole of the rotation axis and the lines show
the stellar equator. Note that $m=-1$ perturbations are $\pi$ out of
phase with the $m=1$ (and have not been plotted here). When the
rotation axis lies in the plane of the sky ($i_{\rm s} = 90^\circ$),
the $m=\pm 1$ components presents their strongest observable
amplitudes. In contrast, the $m=0$ component cannot be detected
because the intensity perturbations in the northern and southern
hemispheres cancel at all phases of the pulsation cycle, giving no
disk-averaged signal. The situation is reversed at $i_{\rm s} =
0^\circ$, when the rotation axis lies along the line-of-sight and
perturbations due to the $m= \pm 1$ components are no longer visible
owing to geometric cancellation.

This dependence (measured in power) may be written explictly as:
 \begin{equation}
 {\cal E}_{lm}(i_{\rm s}) = \frac{(l-|m|)!}{(l+|m|)!}  \left[
   P_l^{|m|} (\cos i_{\rm s}) \right]^2,
 \label{eq:inc}
 \end{equation}
where $P_l^{|m|}$ is the Legendre function, and the sum over ${\cal
  E}_{lm}(i_{\rm s})$ is normalized to unity. Measuring the
relative power of the azimuthal components of different $|m|$ in a
non-radial multiplet therefore provides a direct estimate of the
stellar angle of inclination, $i_{\rm s}$, or more properly $|i_{\rm
  s}|$ since symmetries inherent in Equation~\ref{eq:inc} mean we
cannot discriminate between $i_{\rm s}$ and $-i_{\rm s}$, and $\pi -
i_{\rm s}$ and $\pi + i_{\rm s}$.

The above discussion rests on two assumptions. Firstly, that
contributions to the observed stellar intensity across the visible
stellar disk depend only on the angular distance from the disk centre.
This is valid for photometric observations, where limb darkening
controls the weighting. Secondly, that there is equipartition of
energy between the different $m$ components\footnote{While the case
  for stochastically excited and intrinsically damped solar-like
  oscillations leads to energy equipartition, for observations made
  over a sufficient number of lifetimes, this is not so for classical
  ``heat-engine'' pulsators (e.g., the $\gamma$\,Doradus,
  $\delta$\,Scuti and white-dwarf classes).}. This should be valid
except in very rapid rotators where rotation can affect convection,
which excites and damps the modes. The predicted power asymmetries
(Belkacem et al. 2009) of our stars are of the order of 1\,\%, which
are negligible for our analysis.

Fig.~\ref{fig:example} shows the appearance in the frequency-power
spectrum of an idealized $l=1$ multiplet as a function of the angle
$i_{\rm s}$ (see also Gizon \& Solanki 2003). Panels in the
left-hand column correspond to the cases shown in
Fig.~\ref{fig:spher}.  The $l=1$ modes are approximately three times
more prominent in the frequency-power spectrum than the $l=2$ modes
(see, e.g., Ballot et al.\ 2011b). Hence, it is these modes that
largely constrain our ability to infer $i_{\rm s}$.

The individual components in Fig.~\ref{fig:example}, which are plotted
in blue ($m=\pm 1$) and red ($m=0$), were modelled as Lorentzian
functions, the underlying function used to describe the damped p
modes. The width $\Gamma$ of each Lorentzian -- which is proportional
to the mode damping rate -- is $3.0\,\rm \mu Hz$, which corresponds
approximately to the linewidths observed in the most prominent $l=1$
modes of Kepler-50 and Kepler-65. The splitting is $\delta\nu_{\rm
  s}=1.5\,\rm \mu Hz$, which corresponds to a rotation period of
 $7.7\,\rm d$, and so matches approximately what we observe
for the two stars. The thick black lines show the combined multiplet
profiles.

Given sufficient resolution in frequency and good S/N in the modes, it
is the ratio $\delta\nu_{\rm s}/\Gamma$ of intrinsic stellar
properties that determines whether it is possible to resolve the
components, and hence to infer the true underlying ${\cal
  E}_{lm}(i_{\rm s})$ and hence the value of $i_{\rm s}$. As noted
previously, at angles close to $90^\circ$ the $m=0$ component has
insignificant visibility and the overall appearance is dominated by
the $|m|=1$ components (the converse being true at angles close to
$0^\circ$). Uncertainties in the inferred angle will be largest when
the $i_{\rm s}$ matches these extreme cases, all other factors being
equal (Ballot et al. 2008). This is because there are then only modest
variations in the overall appearance of the mode multiplet with
changing $i_{\rm s}$.

\subsection{Estimation of stellar inclination angles}
\label{sec:anganal}

Extracting  the required information from the rotationally split
components proceeds via a careful fitting of the modes in the observed
frequency-power spectrum, sometimes referred to as peak-bagging (see
Appourchaux et al. 2012, and references therein, for further results
on \emph{Kepler} targets).

\begin{figure*}
\epsscale{1.1}

\plottwo{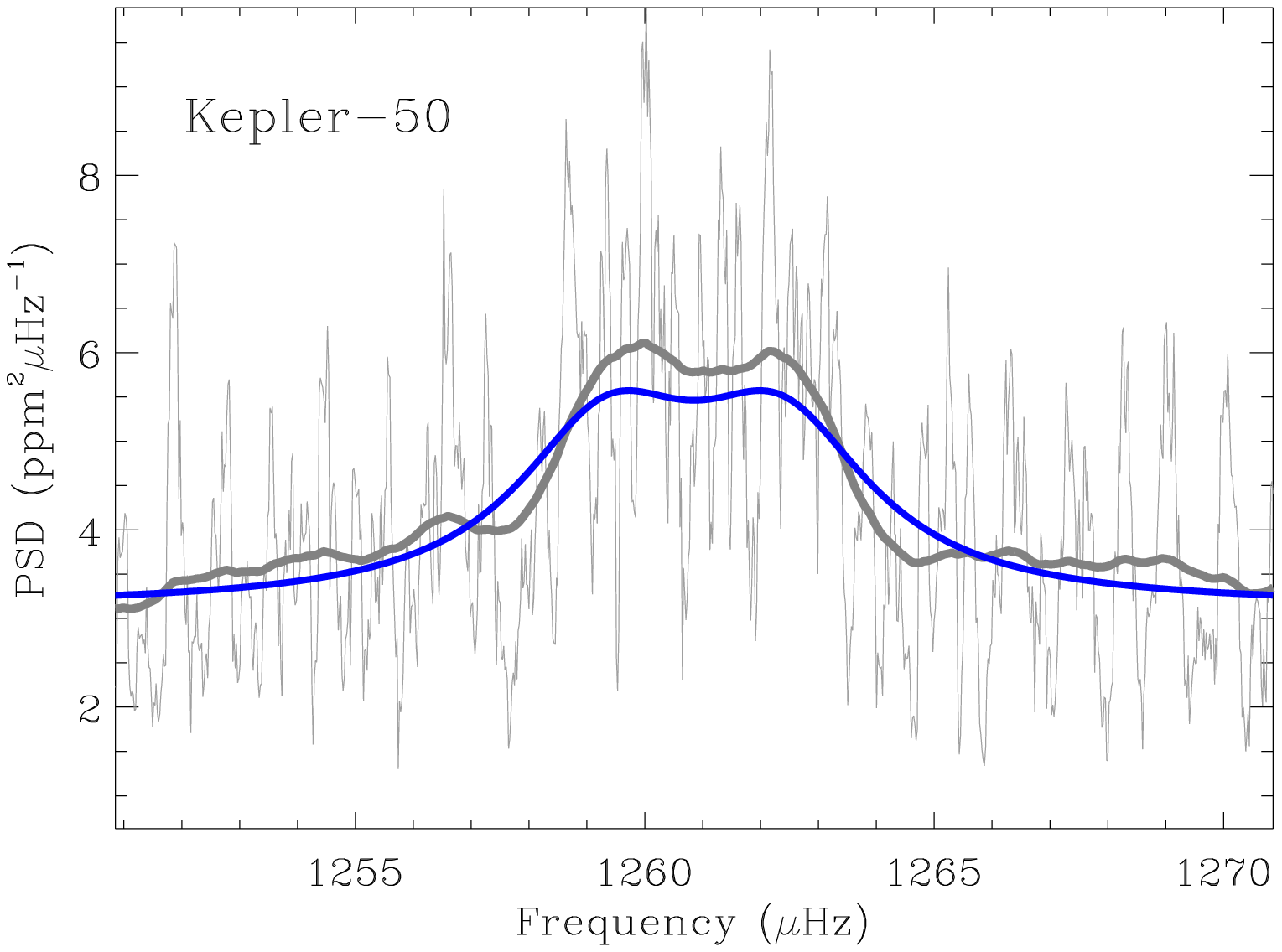}{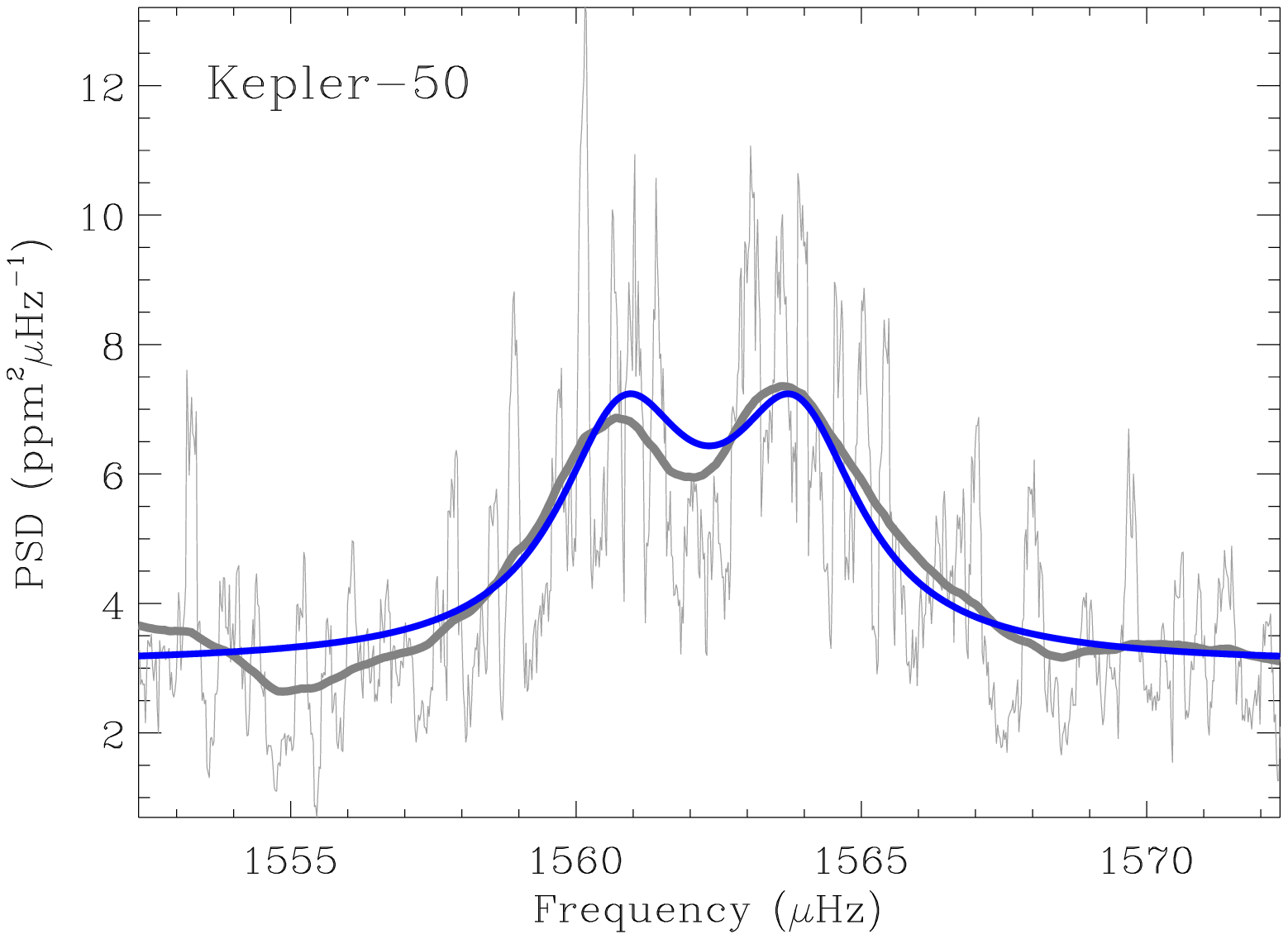}

\plottwo{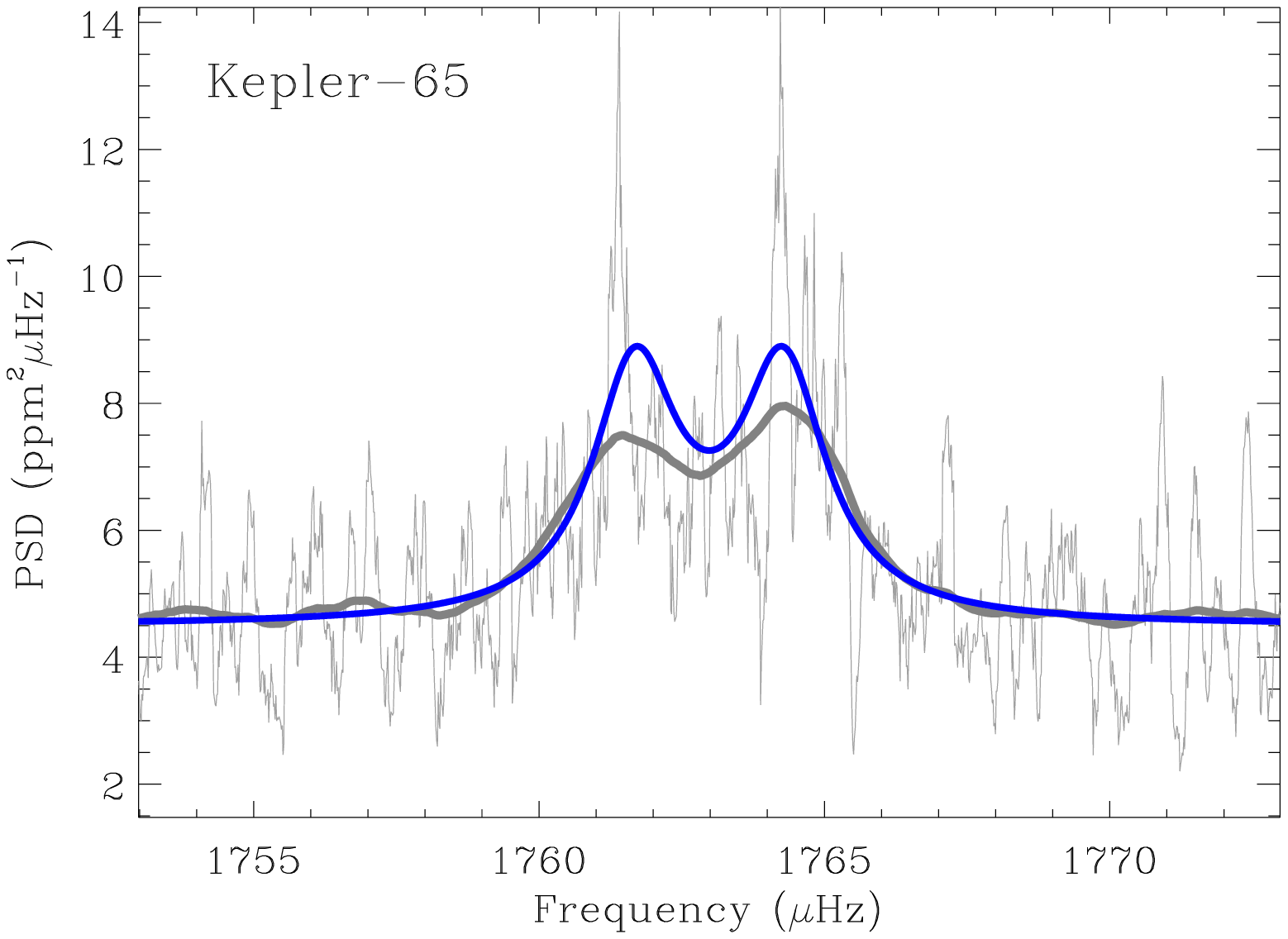}{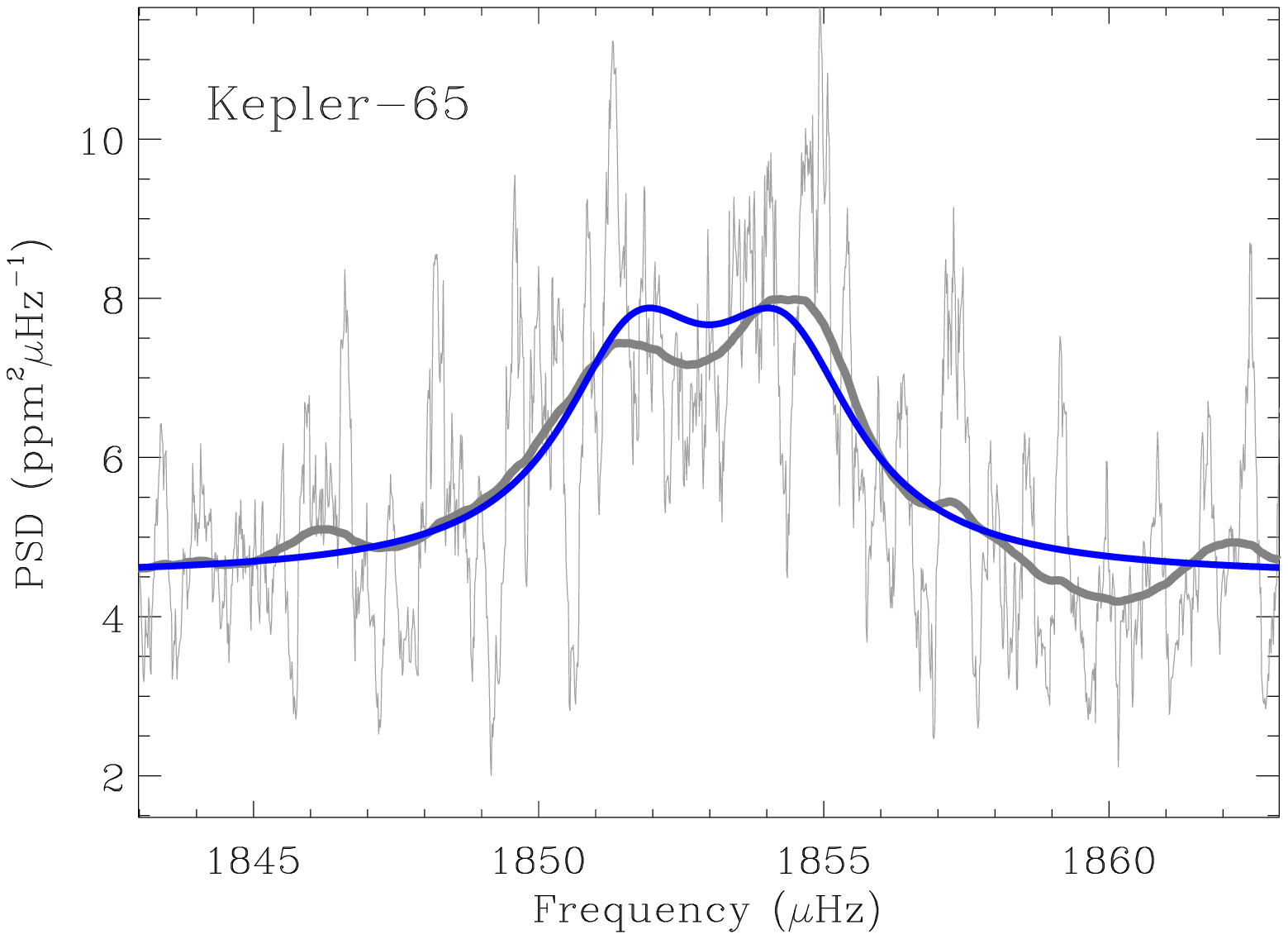}

\caption{Prominent $l=1$ modes in the frequency power spectra of
  Kepler-50 (top panels) and Kepler-65 (bottom panels). Light grey
  lines: observed spectra after applying light smoothing. Thick dark
  grey lines: observed spectra after applying a heavier smoothing of
  width $1.5\,\rm \mu Hz$. Dark-blue lines: best-fitting models from
  MCMC analysis.}

\label{fig:modes}
\end{figure*}

Frequency splittings $\delta\nu_{\rm s}$ due to rotation are clearly
visible in the oscillation spectra of both stars. Fig.~\ref{fig:modes}
shows two prominent $l=1$ modes in each star. The light grey lines
plot the observed spectra after applying a light amount of
smoothing. The thick dark grey lines follow the spectra after they
have been smoothed with a filter of width $1.5\,\rm \mu Hz$, which
provides an approximate representation of the underlying (noise-free)
profiles. Even without a detailed analysis it is apparent that the
observed modes bear a striking resemblance to the high-inclination
cases in Fig.~\ref{fig:example}. The dark-blue lines follow the
best-fitting Lorentzian models, which we describe below.


The observed frequency-power spectrum $\mathcal{P}(\nu)$ of each star
was modelled as
 \begin{equation}
 \mathcal{P}(\nu) = \mathcal{O}(\nu) + B(\nu),
 \label{eq:specmodel1}
 \end{equation}
where $\mathcal{O}(\nu)$ describes the oscillations and $B(\nu)$
contains background terms due to granulation, activity and photon shot
noise.  The oscillations $\mathcal{O}(\nu)$ were modelled as a series
of Lorentzian profiles describing the stochastically excited and
intrinsically damped modes. We adopted a global description, in which
we modelled simultaneously all the observable modes instead of
modelling and analyzing the spectrum one order at a time. This
approach improves the accuracy of the modelling because it takes
proper account of the power from the slowly decaying Lorentzian peaks
that bleeds in frequency between the neighbouring modes.  The
modelled oscillations spectrum was thus described by:
 \begin{equation}
 \mathcal{O}(\nu) = \sum_{n',l} \sum_{m=-l}^{l}
 \frac{{\cal E}_{lm}(i_{\rm s}) H_{n'l} }{1 + 4/\Gamma_{n'}^2 (\nu -
   \nu_{n'l} - m \delta\nu_{\rm{s}})^2},
 \label{eq:specmodel2}
 \end{equation}
The inner sum in the above runs over the $m$ components of each
rotationally split multiplet; while the outer sum runs over all
observed modes, in radial order $n$, and degree $l$. Note that the
dummy variable $n'$, which tags the radial order, takes values $n'=n$
for $l=0$ and $l=1$ modes, and $n'=n-1$ for $l=2$ modes (which lie
adjacent in frequency to $l=0$ modes of $n'=n$). The angle $i_{\rm s}$
and single splitting parameter $\delta\nu_{\rm s}$ are two of the
parameters to be optimized, along with the frequencies $\nu_{nl}$ used
to estimate the fundamental stellar properties
(Section~\ref{sec:prop}).

The parameters $H_{nl}$ and $\Gamma_{n}$ describe the height (maximum
power spectral density) and linewidth of each mode. We fit a single
linewidth parameter to each order. The relative heights of the
components in each non-radial multiplet are controlled by $i_{\rm s}$
through the function ${\cal E}_{lm}(i_{\rm s})$
(Equation~\ref{eq:inc}). The heights are constrained by the relation
$H_{n'l} = H_{n'0} V^2_{l}$, where the parameter $V^2_{l}$ describes
the visibilities of modes of different $l$, relative to $l=0$. The
visibilities are given by integrating the spherical harmonic functions
over the visible disk with suitable allowance made for limb darkening
and the spectral bandpass of the observations (Ballot et
al.\ 2011b). We adopted fixed values of $V^2_0=1.0$, $V^2_1=1.5$ and
$V^2_2=0.5$ in our analysis.

The background was modelled as the sum of three components: a flat
photon shot-noise component, $W$, and two frequency-dependent
components to describe contributions from granulation and
activity. For the latter components, we used functions based on the
Lorentzian-like forms proposed by Harvey (1985), which provide a good
description of the observed backgrounds in solar-type stars (e.g., see
Metcalfe et al. 2012). The composite background was then described by:
 \begin{equation}
 B(\nu) = W + \sum_{k=1}^2\frac{4 \sigma_k^2 \tau_k}{1 + (2\pi\tau_k\nu)^2 + 
         (2\pi\tau_k\nu)^4},
 \label{eq:backmodel1}
 \end{equation}
with $k=1$ associated with the granulation component, and $k=2$
associated with the activity component. The granulation and activity
components each have two free parameters to be optimized: $\sigma_k$
is related to the RMS amplitude of the signal in the time domain,
while $\tau_k$ is the characteristic timescale of the decaying
autocorrelation function. For granulation, $\sigma$ and $\tau$ are
smaller than the corresponding activity parameters.

We adopted two different approaches to the fitting, and hence to
estimation of $i_{\rm s}$.  In the first approach the parameters of
the model in Equation~\ref{eq:specmodel1} were optimized using a MCMC
approach, as described by Handberg \& Campante (2011). We adopted a
flat prior for $i_{\rm s}$ between $0^\circ$ and $90^\circ$. In order
to avoid the rejection of sample jumps close to the boundaries --
i.e., those that would jump beyond the range set by the prior -- in
practice we selected from samples in the range $-90^\circ$ to
$180^\circ$ and modified accepted jumps that went beyond the allowed
range by reflecting about the $i_{\rm s}=0^\circ$ and $90^\circ$
boundaries. A flat prior was imposed on $\delta\nu_{\rm{s}}$, running
between zero and $2.5\,\rm \mu Hz$ for Kepler-50, and zero and $5\,\rm
\mu Hz$ for Kepler-65. Besides making it possible to incorporate
relevant prior information through Bayes' theorem, the MCMC approach
also gave the marginal probability density function (PDF) of each of
the model parameters (e.g., see discussion in Appourchaux 2011). In
order to provide a cross-check we also used Maximum Likelihood
Estimation (MLE) to fit the spectrum (e.g. see Duvall \& Harvey 1986;
Toutain \& Appourchaux 1994), using the so-called pseudo-global
fitting described by Fletcher et al. (2009). Rather than fit the
$i_{\rm s}$ directly with MLE, we instead performed a series of MLE
fits with the angle fixed at different values, the aim being to sample
the maximized likelihood of the best-fitting model as a function of
the chosen $i_{\rm s}$. Even though $i_{\rm s}$ is independent of the
splitting parameter $\delta\nu_{\rm s}$, the measured values are often
highly correlated (e.g., see Ballot et al.\ 2006, 2008). In order to
constrain the two individual parameters, or their combination the
so-called reduced splitting (i.e., $\delta \nu_{\rm s }\,\sin i_{\rm
  s}$), it is desirable to have access to the corresponding maximized
likelihood in two-dimensional parameter space. We therefore performed
fits with both $i_{\rm s}$ and $\delta\nu_{\rm s}$ taking values on a
dense grid. This yielded a two-dimensional grid of maximized
likelihoods, making possible inference on the inclinaton and splitting
from construction of confidence intervals based on the likelihood
surface. The MLE approach had the advantage of being more
computationally efficient than the MCMC analysis. However, given that
the input values for the inclination and splitting are fixed prior to
the fitting, one cannot extract a bona fide posterior probability
distribution.  The MCMC and MLE approaches returned results in
excellent agreement. Here, we present those given by the MCMC
approach.

Table~\ref{tab:resang} lists the final MCMC estimates of the
inclinations and splittings, together with their corresponding
$1\sigma$ credible regions. The estimated $i_{\rm s}$ of both stars
are consistent with $90^\circ$, to within the uncertainties. We note
that the final values for the splittings were given by the median of
each posterior distribution, while for the angles we opted to use the
mode of each distribution. The rationale behind this decision was
twofold. Firstly, the PDF of the inclination is truncated at $i_{\rm
  s} = 90^\circ$ and the median is thus not a representative statistic
of the distribution. Secondly, in each case the model of the
oscillations spectrum built by using the mode for the inclination,
together with the median for all the remaining parameters (including
the splitting) has a higher posterior likelihood than the models made
using exclusively either the median or the mode for all parameters.


\begin{deluxetable}{cccc}
\tabletypesize{\scriptsize} 
\tablecaption{Estimated stellar inclinations and rotational splittings} 
\tablewidth{0pt}
\tablehead{
 \colhead{Star}& \colhead{$i_{\rm s}$}& \colhead{$\sin\,i$}& \colhead{$\delta\nu_{\rm s} \equiv \Omega/2\pi$}\\
 \colhead{}& \colhead{(degrees)}& \colhead{ }& \colhead{($\rm \mu Hz$)}}
\startdata
                &              &                   &                    \\            
Kepler-50&  $82^{+8}_{-7}$& $0.99^{+0.01}_{-0.02}$& $1.51^{+0.09}_{-0.08}$\\
                &              &                   &                    \\            
Kepler-65& $81^{+9}_{-16}$& $0.99^{+0.01}_{-0.08}$& $1.30^{+0.19}_{-0.16}$\\
\enddata
\label{tab:resang}
\end{deluxetable}

\begin{figure*}
\epsscale{1.0}
\plotone{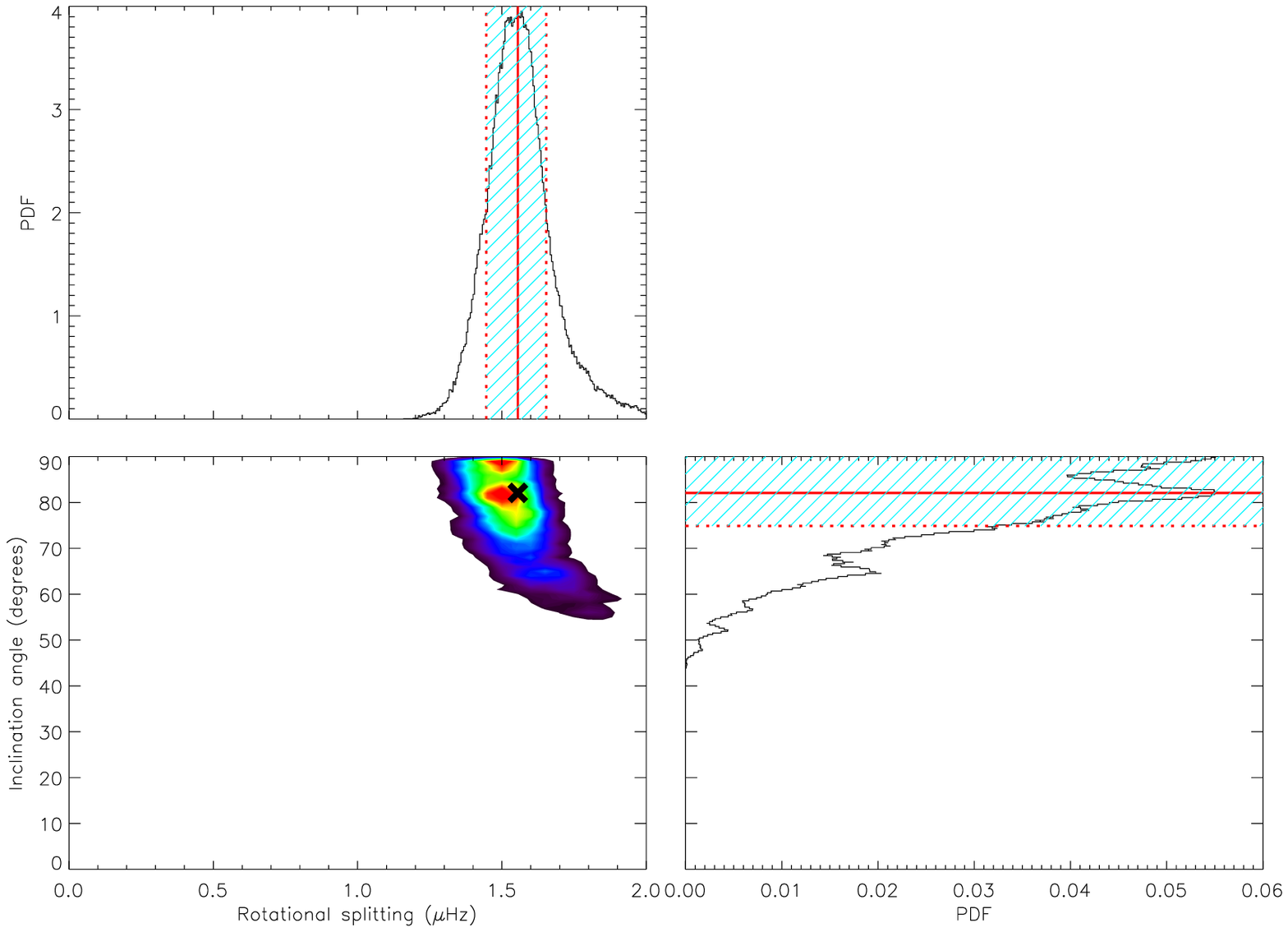}

\caption{Asteroseismic results on Kepler-50. Lower left-hand panel:
  Correlation map in angle of inclination $i_{\rm s}$ and rotational
  frequency splitting $\delta\nu_{\rm s}$ (highest likelihoods
  rendered in red); Top and right-hand panels: PDFs obtained after
  marginalization. Note that the PDFs are normalized so that the
  integral under each curve is unity. Bold crosses mark the final
  parameter estimates given in Table~\ref{tab:resang}.}

\label{fig:corrmap262}
\end{figure*}
\begin{figure*}
\epsscale{1.0}
\plotone{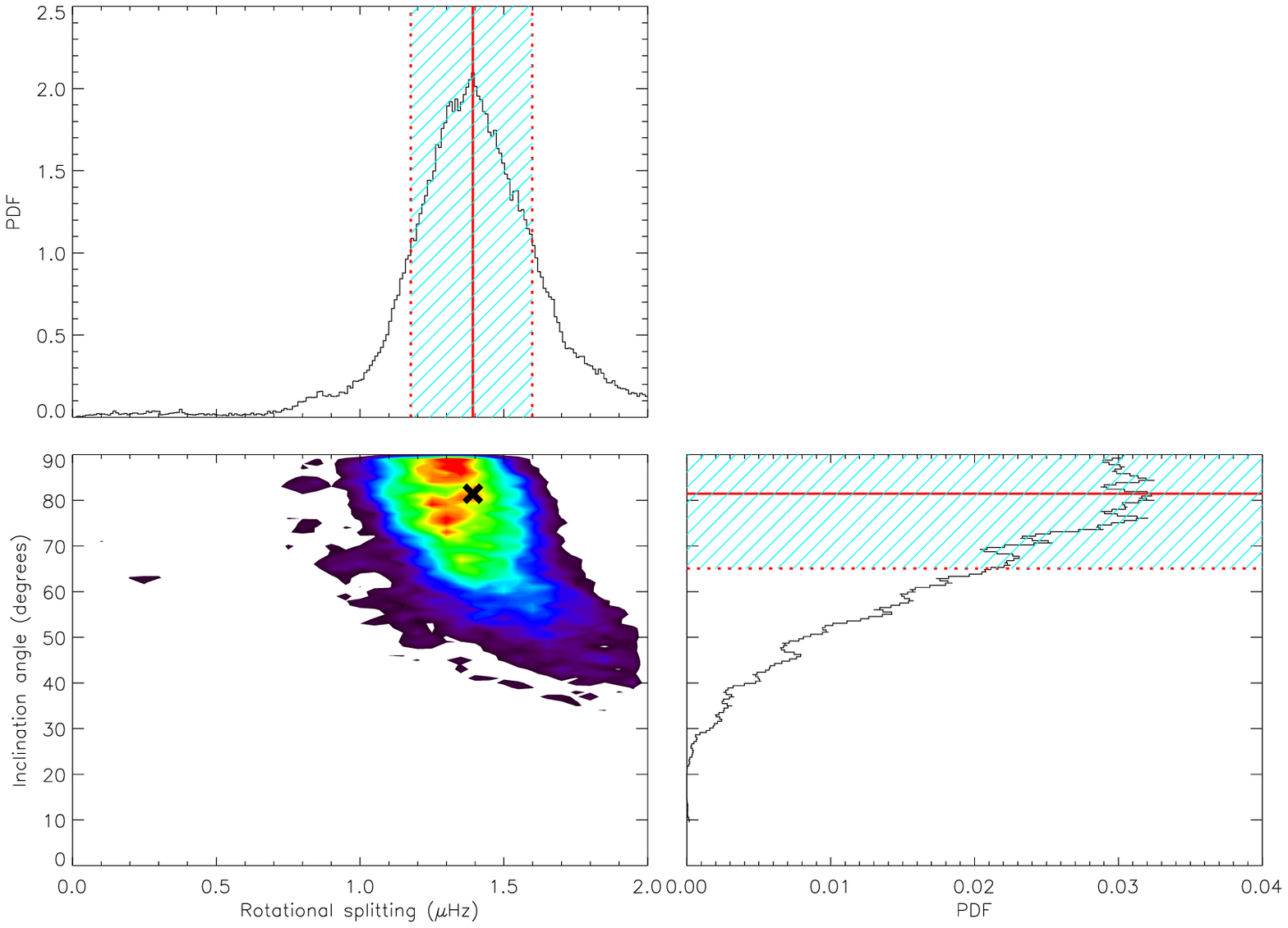}

\caption{Similar to Fig.~\ref{fig:corrmap262}, but for Kepler-65.}

\label{fig:corrmap85}
\end{figure*}

As noted above, the thick black lines in Fig.~\ref{fig:modes} show
best-fitting models from the MCMC analysis across frequency ranges
occupied by two $l=1$ modes in each star. Figs.~\ref{fig:corrmap262}
and~\ref{fig:corrmap85} show the correlation maps in the angle and
splitting, as well as the PDFs obtained after
marginalization. Binwidths for the PDFs were fixed by following the
procedure given in Scott (1979) [see also Handberg \& Campante 2011].


\section{Comparison with measures of surface rotation}
\label{sec:surfrot}

We have compared the asteroseismic results from
Section~\ref{sec:anganal} with two independent estimates of the
surface rotation: one extracted from signatures of rotational
modulation in the \emph{Kepler} lightcurves, and another extracted
from spectroscopic data on both stars.

If a star has spots on its surface then rotation will carry the spots
in and out of view, inducing quasi-periodic flux variations. Such
variations have been detected for many stars, and it is not unusual to
see activity in stars as hot as our host stars (Basri et
al.\ 2011). The rotation period of Kepler-50 has already been detected
in \emph{Kepler} data (Hirano et al.\ 2012a) and Kepler-65 also shows
clear signs of rotational modulation in its lightcurve.

\begin{figure*}
\epsscale{0.9}
\plotone{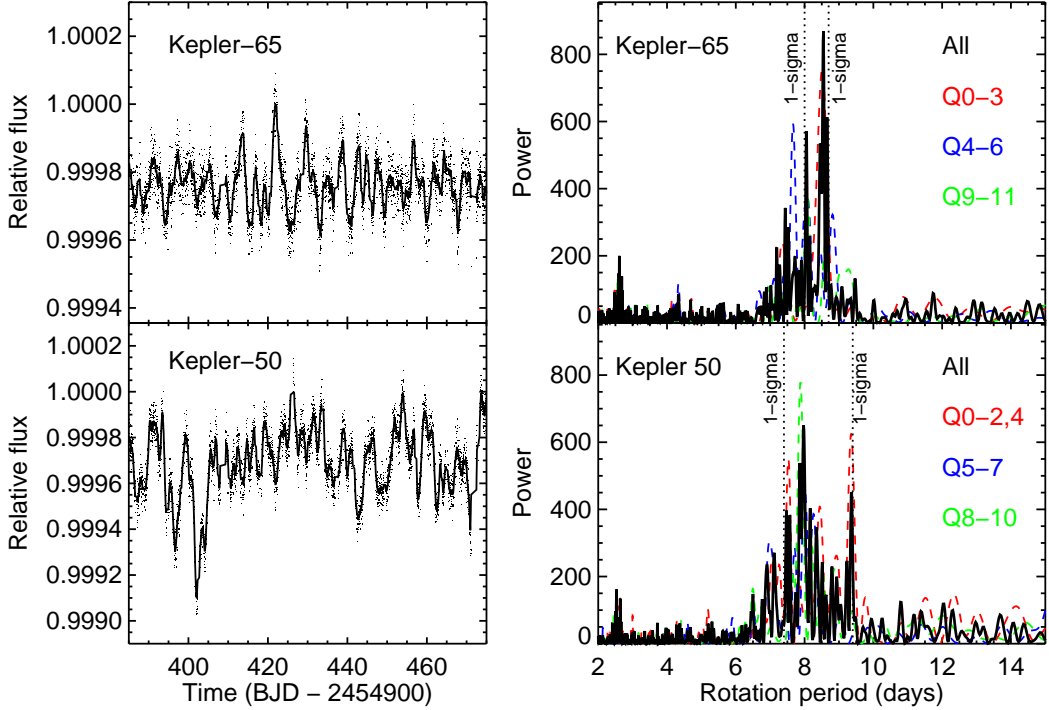}

 \caption{Left-hand panels: Three-month segments of the long-cadence
   PDC-MAP data (from Q5), in which intrinsic stellar variability on
   time scales of days is evident. The dots show the de-trended data
   (see text), and the thick line represents a smoothed version (10-hr
   boxcar). Right-hand panels: Lomb-Scargle periodograms of the
   PDC-MAP data of both stars. Thick black lines: periodograms from
   using all corrected data.  Dashed, colored lines: periodograms of
   three independent three-quarter-long segments of data. Confidence
   intervals on the quoted average periods are marked by the vertical
   dotted lines.}

 \label{fig:surf} 
\end{figure*}

The raw \emph{Kepler} data are known to suffer from systematic trends
that appear to be shared by most of the stars on a given CCD detector
module. The absolute effect of these trends on the measured stellar
fluxes is much larger than the activity levels for both stars, so we
needed to choose an appropriate detrending algorithm that would
suppress the unwanted systematics without removing the astrophysical
signal. One such algorithm available to us is PDC-MAP (Stumpe et
al.\ 2012, Smith et al.\ 2012), which was developed by the
\emph{Kepler} team. Principal component analysis is used to extract a
basis of co-trending vectors that capture the systematic trends in
each module. Each stellar flux datum may be decomposed into a linear
combination of the astrophysical variability and the co-trending
vectors. One could perform a least-squares fit to extract those
coefficients but PDC-MAP goes one step further. During a first pass on
the data it applies a least-squares approach to obtain the
coefficients for all stars that fall on a given detector module. The
distributions are then used as priors in a Bayesian sense, thereby
mitigating possible over-correction of the time series for any
individual star.  PDC-MAP appears to do a reasonable job of
eliminating the systematic trends in almost every quarter of data, but
does fail on a few occasions. From the available long-cadence (LC)
data (Jenkins et al. 2010) collected through Q11, we discarded the Q3
and Q11 data for Kepler-50, and the Q7 and Q8 data for Kepler-65.
Outliers were also removed by performing 3$\sigma$ clipping using a
10-hour-long moving-median filter. The remaining data were quite
adequate to estimate robust rotation periods. The left-hand panels of
Fig.~\ref{fig:surf} show three-month segments of the data, in which
intrinsic stellar variability on time scales of days is evident. This
variation is smaller than that displayed by the active Sun (e.g., see
Basri et al. 2010, 2011), for which the semi-amplitude of the
variability reaches levels close to 1 part in $10^3$.

To extract estimates of the surface rotation periods we followed the
analysis described by Hirano et al.\ (2012a). We first calculated the
Lomb-Scargle periodogram of each set of processed data, which are
plotted in the right-hand panels of Fig.~\ref{fig:surf}. Both stars
show significant peaks around 8\,days due to rotational modulation of
spots. We checked that our results did not depend on the detrending,
as follows. Firstly, we used the simple least-squares fit to the
co-trending vectors described above. Secondly, we applied a 20-day
median smoothing filter to divide the raw data.  Both approaches led
to periodograms in good agreement with our main results.

Both stars show several significant, closely spaced peaks in their
periodograms. The spread in period of the peaks will have
contributions from the finite spot lifetimes and might also suggest
the presence of surface differential rotation. Fig.~\ref{fig:surf}
also shows periodograms obtained from particular subsets of the data,
which suggest that the rotation periods might be evolving with
time. Again, this might be associated with stochastic variability due
to the spot lifetimes, or it could have a contribution due to changes
in the spot latitudes.  These signatures are not unexpected for such
hot stars (Collier Cameron 2007) but the low S/N of the stellar
variability in the lightcurves makes it hard to test the results in
greater detail. It is worth adding that we also checked that the
spread of significant periods was not an artifact of the observational
window function. We sampled commensurate sine waves of period 8\,d at
the same time stamps as the real, cleaned PDC-MAP lightcurves, and
found that the resulting periodograms displayed a much smaller width
of significant periods than the real data (of order 0.2 to 0.4\,d).

As in Hirano et al.\ (2012a), for each star the continuous range of
periods containing all peaks with more than half the power of the
highest peak was taken as the uncertainty on the surface rotation
period, with the center of the range adopted as the quoted period,
$P_{\rm rot}$. The rotation periods obtained in this way were $8.4 \pm
1.0\,\rm days$ for Kepler-50 and $8.4 \pm 0.3\,\rm days$ for
Kepler-65. The rotational frequency splittings $\delta\nu_{\rm s}$
given by the asteroseismic analysis are equivalent to rotation periods
$P_{\rm rot} = 1/\delta\nu_{\rm s}$ of $7.7^{+0.5}_{-0.4}\,\rm days$
for Kepler-50 and $8.9^{+1.3}_{-1.1}\,\rm days$ for Kepler-65. The
close agreement between the surface and internal rotation rates is
consistent with the expectation that the frequency splittings of p
modes observed in main-sequence stars are largely determined by the
rotation profile in the stellar envelope (as in the case of the Sun).

We can also compare our results with $v\sin\,i_{\rm s}$ estimates
extracted from the spectroscopic data (see Section~\ref{sec:prop}).
For Kepler-50 the spectroscopic analyses gave $8.6 \pm 0.8\,\rm
km\,s^{-1}$ (SME) and $10.3 \pm 0.5\,\rm km\,s^{-1}$ (SPC), while for
Kepler-65 they gave $9.8 \pm 0.8\,\rm km\,s^{-1}$ (SME) and $11.9 \pm
0.5\,\rm km\,s^{-1}$ (SPC). To compare with the asteroseismic results,
we converted the estimated rotational frequency splittings to
projected rotational velocities using
 \begin{equation}
 v \sin \,i_{\rm s} \equiv 2\pi R\,\delta\nu_{\rm s}\sin\,i_{\rm s},
 \label{eq:vsini}
 \end{equation}
with the stellar radii $R$ given by the asteroseismic analysis
discussed in Section~\ref{sec:prop}. These conversions gave equivalent
asteroseismically determined projected velocities of
$8.0^{+1.2}_{-1.0}\,\rm km\,s^{-1}$ for Kepler-50 and $10.4 \pm
0.6\,\rm km\,s^{-1}$ for Kepler-65. Again, we find agreement between
the asteroseismic and surface estimates. We note that the SPC
$v\sin\,i_{\rm s}$ values are higher than the SME values by about
20\,\%, in agreement with the findings by Torres et al.\ (2012).

Finally, we may combine the spot modulation and $v\sin\,i_{\rm s}$
results to provide independent estimates of $i_{\rm s}$, via
 \begin{equation}
 \sin\,i_{\rm s} = P_{\rm rot} (v\sin\,i_{\rm s})/(2\pi R). 
 \label{eq:sini}
 \end{equation}
When we used the SME results for $v\sin\,i_{\rm s}$ we found that
sines of the angles were constrained at the $1\sigma$ level to lie
above 0.89 for Kepler-50 and 0.90 for Kepler-65, again implying that
both stars have their rotation axes nearly perpendicular to the line
of sight.  When the SPC $v\sin\,i_{\rm s}$ were used we obtained
$\sin\,i_{\rm s} > 1.0$ for both systems suggesting that (at least for
these systems) the SPC results are overestimated.


\section{Discussion}
\label{sec:disc}

Our central result is that the host stars of the Kepler- 50 and
Kepler-65 planetary systems have their rotation axes nearly
perpendicular to the line of sight, with $\sin i_{\rm s}$ constrained
at the $1\sigma$ level to lie above 0.97 and 0.91, respectively.
Expressed in terms of angles, we have $|90^\circ - i_{\rm s}| <
15^\circ$ for Kepler-50 and $<25^\circ$ for Kepler-65. The orbital
inclinations of the planets in each system are also near 90$^\circ$,
with a deviation of only $\approx 5^\circ$ for the planets of
Kepler-50 and $\approx 2^\circ$ for the planets of
Kepler-65. Therefore our observations are consistent with small
differences in the stellar and orbital inclination angles, and low
stellar obliquities.

A limitation of the results is that it is possible for the obliquity
to be large but for the difference in inclination angles to be small.
Spherical geometry dictates that the 3D obliquity angle $\psi$
between the stellar spin axis and the planetary orbital axis is given
by
\begin{equation}
\cos\psi = \cos i_{\rm s} \cos i_{\rm p} + 
\sin i_{\rm s} \sin i_{\rm p} \cos \lambda,
\end{equation}
where $\lambda$ is the projected angle on the sky between the orbital
and rotational angular momentum vectors, which the asteroseismic
method cannot provide. This is the converse of the situation with, for
example, the RM and spot-occultation methods, for which the data
reveal $\lambda$ but are generally insensitive to $i_{\rm s}$. To
overcome this limitation, one would need to combine the various
measurement techniques to determine both $\lambda$ and $i_{\rm s}$ for
the same system, or else conduct observations of many systems and
perform a statistical analysis of the ensemble (see, e.g., Fabrycky \&
Winn 2009). Measuring the RM effects given by the planets in the two
systems considered in this paper would be very challenging, due to the
small sizes of the planets. The highest-amplitude RM signal for
Kepler-50 would be around $\simeq 0.3\,\rm m\,s^{-1}$ (given by
Kepler-50c), while for Kepler-65 it would be $\simeq 1.8\,\rm
m\,s^{-1}$ (given by Kepler-65c).

While the two stars we have analyzed are special -- in the sense that
they have transits -- it is important to test the null hypothesis that
they are members of a population of stars randomly oriented in space.
Put another way: had the stars been randomly oriented, what is the
chance that the inclination angles would have been as close to
$90^\circ$, as observed? For an isotropic distribution, the
probability of observing $|90^\circ - i_{\rm s}| < x$ is simply $\sin
x$. Evaluating this for Kepler-50 and Kepler-65 we find the chance to
be 26\,\% and 42\,\% individually, and the chance of observing both of
them so close to $90^\circ$ is 11\,\%.  In this sense there is only an
11\,\% chance we would have obtained our results in the absence of any
correlation between the orientations of the stellar rotation and
planetary orbits.

Despite these limitations, let us consider the implications of low
obliquities. The number of extrasolar multiplanet systems for which
the stellar obliquity has been measured is now four, with Kepler-50
and Kepler-65 joining the previously studied systems Kepler-30
(Sanchis-Ojeda et al.\ 2012) and KOI-94 (Hirano et al.\ 2012b).  In
all four cases, the obliquity is consistent with zero. We have already
summarized our results for Kepler-50 and Kepler-65, and for the other
two systems $\lambda$ was found to be consistent with zero to within
about $10^\circ$.

Several years ago these results would have seemed mundane. Low
obliquities are expected in the standard picture in which the star and
planets have the same direction of angular momentum originating from a
common accretion disk. Furthermore, up until 2008, low obliquities had
been observed in all the exoplanetary systems that had been examined
(all involving single hot Jupiters). Since that time, systems with hot
Jupiters have been found to have host stars with a wide range of
obliquities (Albrecht et al.\ 2012). Those results have been taken as
evidence that the process that produces hot Jupiters also tilts their
orbits relative to the initial plane of their formation. Specifically,
the results have been taken to support theories for the origin of hot
Jupiters involving few-body dynamics and tidal circularization
(alternatively referred to as high-eccentricity migration) as opposed
to the formerly dominant paradigm of disk migration.

It has been pointed out, however, that this conclusion depends
critically on the assumption that the current stellar equatorial plane
is aligned with the original plane of the planetary orbits. Several
papers have questioned this assumption. Bate et al.\ (2010) suggested
that the chaotic environment of a star-forming region might lead to
large mismatches between the direction of stellar rotation and the
orientation of the late-stage protoplanetary disk. Thies et
al.\ (2011) proposed that inclined planets arise from capture of gas
from a neighboring star.  Lai et al.\ (2011) presented a theory of
magnetic interaction between a young star and the inner edge of its
accretion disk that can tip the star by a significant angle. Rogers et
al.\ (2012) considered stars with convective cores and radiative
envelopes, and found that they might be susceptible to a directional
wandering of the outermost layers of the star due to transport of
angular momentum by internal gravity waves from the
convective-radiative boundary.

In a few cases, it has been possible to compare the orientation of
stellar rotation and the orientation of a surrounding disk (Le Bouquin
et al.\ 2009, Watson et al.\ 2011), and the results have favored the
hypothesis of close alignment and low obliquities. However, more
stringent tests are warranted, and are provided by the results for the
four multiplanet systems Kepler-30, 50, 65, and KOI-94. The low
obliquities suggest that high obliquities are confined to the
hot-Jupiter systems, or at least have not yet falsified that
hypothesis. The results are therefore starting to provide support to
the argument that most or all hot Jupiters are formed through
inclination-lifting processes and not via disk migration.

All four of the stars are cool enough to have outer convective zones
(indeed, Kepler-50 and Kepler-65 must have convection because they
exhibit solar-like oscillations). It is unclear how effective the
model of Rogers et al.\ (2012) might be for stars with thin convective
envelopes (e.g., like Kepler-50, Kepler-65 and KOI-94, which are all
hotter than the Sun).  It is worth noting that this theory already has
difficulty accommodating the hot-Jupiter results involving cool,
convective stars with high obliquities, namely HAT-P-11 (Winn et
al.\ 2010b, Sanchis-Ojeda \& Winn 2011), HD~80606 (Winn et al.\ 2009,
Pont et al.\ 2009, H\'ebrard et al.\ 2010), and WASP-8 (Queloz et
al.\ 2010).

Of course, one should not be satisfied with a sample of only four
systems, especially given the limitations caused by projection effects
as noted above. The asteroseismic technique that was deployed in this
work has the advantage that the detectability of the signal is chiefly
a function of the stellar properties, as opposed to the planetary
properties, and therefore has no intrinsic difficulty with small
planets or long-period planets. We expect it will be possible to apply
this technique to a sample of at least 10 other \emph{Kepler} systems
in the near future. It will be possible to draw stronger conclusions
with these results in hand. Definitive conclusions will also be
possible on individual systems with transiting exoplanets when
asteroseismology demonstrates that $i_{\rm s}$ is significantly
different from $90^\circ$.

\acknowledgements Funding for this Discovery mission is provided by
NASA's Science Mission Directorate. The authors wish to thank the
entire \emph{Kepler} team, without whom these results would not be
possible.  W.J.C., T.L.C., G.R.D., Y.E. and A.M.\ acknowledge the
support of the UK Science and Technology Facilities Council
(STFC). J.N.W.\ was supported by the NASA Kepler Participating
Scientist program through grant NNX12AC76G. S.B. acknowledges NSF
grant AST 1105930. Funding for the Stellar Astrophysics Centre (SAC)
is provided by The Danish National Research Foundation. The research
is supported by the ASTERISK project (ASTERoseismic Investigations
with SONG and Kepler) funded by the European Research Council (Grant
agreement no.: 267864).  SH acknowledges financial support from the
Netherlands Organisation for Scientific Research (NWO). Computational
time on Kraken at the National Institute of Computational Sciences was
provided through NSF TeraGrid allocation TG-AST090107. We acknowledge
the Pale Blue Dot Project, hosted by White Dwarf Research Corporation
(whitedwarf.org/palebluedot), and we are also grateful for support
from the International Space Science Institute (ISSI).

\appendix

\section{Estimation of stellar properties using oscillation frequencies}
\label{sec:est}

In the second stage of the stellar properties estimation
(Section~\ref{sec:prop}) three members of the team (SB, JCD and TM)
performed a detailed modelling of the stars using estimates of the
individual oscillation frequencies and the revised spectroscopic data
as inputs.

SB made use of the Yale stellar evolution code, YREC (Demarque et
al. 2008) to model both stars. The input physics included the OPAL
equation of state tables of Rogers \& Nayfonov (2002), and OPAL
high-temperature opacities (Iglesias \& Rogers 1996) supplemented with
low -temperature opacities from Ferguson et al.~(2005).  All nuclear
reaction rates were from Adelberger et al.~(1998), except for the rate
of the $^{14}{\rm N}(p,\gamma)^{15}{\rm O}$ reaction, which was fixed
at the value of Formicola et al.~(2004). Models were constructed for
two values of core overshoot, 0 and 0.2$H_p$. Two families of models
were constructed, one that included the diffusion and settling of
helium and heavy elements as per the formulation of Thoul et
al.~(1994), and one that did not include any diffusion and settling.

YREC was used in an iterative mode whereby the final $T_{\rm eff}$ and
radius for a star of a given mass and metallicity was specified, and
for a given mixing length parameter $\alpha$ the code iterated over
the initial helium abundance $Y_0$ until a model with the specified
$T_{\rm eff}$ and radius was found.  This is similar to the
construction of a standard solar model, although in the solar case
iterations are made over both the mixing length parameter and $Y_0$
with solar age a fixed independent constraint.  Since the ages of the
\emph{Kepler} stars are not known independently, iteration over $Y_0$
were performed for many different values of the mixing length
parameter.  All solutions for which the initial helium abundance was
less than the primordial helium abundance, $Y_{\rm p}$ were
rejected. $Y_{\rm p}$ was assumed to be $0.245$.

Corrections for near-surface effects (the so-called surface term) were
handled in the following manner.  The first step was the construction
of a standard solar model with exactly the same physics as that used
to model the \emph{Kepler} stars. This yielded the set $\nu_{nl\odot}$
of solar model frequencies. These were then used to estimate a set of
``surface term'' frequency offsets, $\delta\nu_{nl\odot}$, for the Sun
by computing differences between the solar model frequencies and the
solar low-degree frequencies observed by the Birmingham Solar
Oscillations Network (BiSON) (as listed in Basu et al.~2009).

For each stellar model, ${\cal M'}$, $\nu_{nl\odot}$ and
$\delta\nu_{nl\odot}$ were then scaled to the mass and radius of
${\cal M'}$ using the homology scaling $r = \left< \Delta\nu({\cal
  M'}) \right> / \left< \Delta\nu_{nl,\odot} \right>$. The resulting
$r\nu_{nl\odot}$-$r\delta\nu_{nl\odot}$ relation was then used to
correct the stellar model for the surface term. Using a least squares
minimization a factor $\beta$ was selected so as to minimize $\sum
\left(\nu^{\rm obs}_{nl}-\nu^{\rm corr}_{nl} \right)^2/
\left(\sigma^{\rm obs}_{nl} \right)^2$ over all observed modes, where
$\nu^{\rm corr}_{nl}=\nu_{nl}^{\cal M'}+ \beta\;r\delta\nu_{nl\odot}$,
with $r\delta\nu_{nl\odot}$ evaluated at $r\nu_{nl\odot}=\nu^{\rm obs}_{nl}$.

JCD followed a prescription that has previously been applied to the
Hubble observations of HD 17156 (Gilliland et al., 2011), and several
\emph{Kepler} exoplanet host stars, i.e.  HAT-P-7
(Christensen-Dalsgaard et al., 2010), \emph{Kepler}-10 (Batalha et
al., 2011) and \emph{Kepler}-36 (Carter et al. 2012). Stellar
evolutionary models were computed with the ASTEC code
(Christensen-Dalsgaard, 2008a).  The calculations used the OPAL
equation of state tables (see Rogers et al. 1996) and OPAL opacities
at temperatures above $10^4\,\rm K$ (Iglesias \& Rogers 1996); at
lower temperature the Ferguson et al. (2005) opacities were
used. Nuclear reactions were calculated using the NACRE parameters
(Angulo et al., 1999). Convection was treated using the B\"ohm-Vitense
(1958) mixing-length formulation. Frequencies were computed for the
models using ADIPLS (Christensen-Dalsgaard, 2008b) and then corrected
for surface effects following the prescription of Kjeldsen et
al. (2008).

For each evolutionary sequence in the grid of ASTEC models, the model
${\cal M}'_{\rm min}$ whose surface-corrected frequencies provided the
best $\chi^2$ match to the observations was selected. The best match
was obtained from application of homology scaling, under the
assumption that in the vicinity of ${\cal M}'_{\rm min}$ frequencies
could be calculated using $r\nu_{nl}({\cal M}'_{\rm min})$, where
$r=[R/R({\cal M}'_{\rm min})]^{-1.5}$, $R$ being the surface radius of
the model. A best-fitting model was then determined by minimizing the
sum $\sum \left( \nu^{\rm obs}_{nl} - r\nu_{nl}({\cal M}'_{\rm min})
\right)^2/ \left(\sigma^{\rm obs}_{nl} \right)^2$ over all observed
modes, as a function of $r$. The resulting minimum value of $r$
defined an estimate of the radius of the best-fitting model along the
given sequence. The other properties of this best-fitting model were
determined by linear interpolation in $R$, to the minimum of $R$.
Statistical analysis of the ensemble of best-fitting properties from
all evolutionary sequences then yielded the final stellar properties,
and their uncertainties (see Christensen-Dalsgaard et al.\ 2010,
Carter et al.\ 2012).

TM used the Asteroseismic Modeling Portal (AMP), a web-based tool tied
to TeraGrid computing resources that uses a parallel genetic algorithm
(Metcalfe \& Charbonneau 2003) to optimize, in an automated manner,
the match to observational data (see Metcalfe et al. 2009, Woitaszek
et al. 2009 for more details). AMP employs the Aarhus stellar
evolution code ASTEC (Christensen-Dalsgaard 2008a) and adiabatic
pulsation code ADIPLS (Christensen-Dalsgaard 2008b).  Models were made
using the OPAL 2005 equation of state and the most recent OPAL
opacities supplemented by Ferguson et al. (2005) opacities at low
temperature, nuclear reaction rates from NACRE (Angulo et al. 1999),
and helium diffusion and settling following Michaud \& Proffitt
(1993). Convection was treated with standard mixing-length theory
without overshooting (B{\"o}hm-Vitense 1958).

Each AMP model evaluation involved the computation of a stellar
evolution track from the zero-age main sequence (ZAMS) through a
mass-dependent number of internal time steps, terminating prior to the
beginning of the red giant stage. The asteroseismic age was optimized
along each evolutionary track using a binary decision tree under the
assumption that $\left< \Delta\nu \right>$ is a monotonically
decreasing function of age (see Metcalfe et al. 2009, and references
therein). The Kjeldsen et al. (2008) prescription was again applied in
an attempt to deal with the surface term.  The optimal model was then
subjected to a local analysis that uses singular value decomposition
(SVD) to quantify the uncertainties of the final model parameters (see
Creevey et al. 2007).

\end{document}